\documentclass[review,10pt]{elsarticle} 
\usepackage{float}
\usepackage{booktabs}
\usepackage{multirow}
\usepackage{geometry}
\usepackage{amsmath}
\usepackage{graphicx}
\usepackage[unicode=true, bookmarks=false]{hyperref}
\usepackage[utf8]{inputenc}
\usepackage{times} 

\usepackage{caption}
\usepackage{amssymb}

\usepackage{tabularray}
\UseTblrLibrary{booktabs}
\usepackage{mathtools}
\usepackage{hyperref}
\usepackage{threeparttable}

\usepackage[superscript]{cite}
\usepackage{xcolor}

\renewenvironment{abstract}{\global\setbox\absbox=\vbox\bgroup
  \hsize=\textwidth%
  \noindent\unskip\textbf{\large Summary}
  \par\medskip\noindent\unskip\ignorespaces}{\egroup}
\makeatletter
\renewcommand\@biblabel[1]{#1} 
\makeatother

\usepackage{tabularx}

\begin{document}
\title{INSIGHT: Spatially resolved survival modelling from routine histology crosslinked with molecular profiling reveals prognostic epithelial-immune axes in stage II/III colorectal cancer}

\author[aff1]{Piotr Keller}
\author[aff1]{Mark Eastwood}
\author[aff2]{Zedong Hu}
\author[aff2]{Aimée Selten}
\author[aff1]{Ruqayya Awan}
\author[aff2,aff3]{Gertjan Rasschaert}
\author[aff2]{Sara Verbandt}
\author[aff9]{Vlad Popovici}
\author[aff4]{Hubert Piessevaux}
\author[aff7]{Hayley T Morris}
\author[aff8]{Petros Tsantoulis}
\author[aff10]{Thomas Alexander McKee}
\author[aff2,aff5]{André D'Hoore}
\author[aff5]{Cédric Schraepen}
\author[aff6]{Xavier Sagaert}
\author[aff6]{Gert De Hertogh}
\author[aff2,aff3]{Sabine Tejpar\corref{cor1}}
\author[aff1]{Fayyaz Minhas\corref{cor1}}

\cortext[cor1]{Joint Supervising Authors}

\affiliation[aff1]{
  organization={Predictive Systems in Biomedicine (PRISM) Lab, Department of Computer Science, University of Warwick},
  city={Coventry},
  postcode={CV4 7AL},
  country={United Kingdom}
}

\affiliation[aff2]{
  organization={Laboratory for Digestive Oncology, KU Leuven},
  city={Leuven},
  country={Belgium}
}

\affiliation[aff3]{
  organization={Gastrointestinal Oncology Department, University Hospitals Leuven},
  city={Leuven},
  country={Belgium}
}

\affiliation[aff4]{
  organization={Gastroenterology Department, Cliniques Universitaires Saint-Luc},
  city={Brussels},
  country={Belgium}
}

\affiliation[aff5]{
  organization={Abdominal Surgery Department, University Hospitals Leuven},
  city={Leuven},
  country={Belgium}
}

\affiliation[aff6]{
  organization={Pathology Department, University Hospitals Leuven},
  city={Leuven},
  country={Belgium}
}

\affiliation[aff7]{
  organization={Department of Pathology, University Hospital Crosshouse; School of Cancer Sciences, University of Glasgow},
  city={Kilmarnock},
  postcode={KA2 0BE},
  country={United Kingdom}
}

\affiliation[aff8]{
  organization={Université de Genève, Faculté de Médecine},
  city={Geneva},
  country={Switzerland}
}

\affiliation[aff9]{
  organization={RECETOX, Masaryk University},
  address={Kamenice 753/5},
  postcode={625 00 Brno},
  country={Czech Republic}
}

\affiliation[aff10]{
  organization={Hôpitaux Universitaires de Genève, Service de Pathologie Clinique; current address: Aurigen SA, Lausanne, Sonic Healthcare Ltd subsidiary},
  city={Geneva / Lausanne},
  country={Switzerland}
}

\begin{abstract}
\textcolor{black}{Accurate prognostication in stage II/III colorectal cancer remains limited, with substantial outcome heterogeneity and unclear links between histology and underlying biology. We present INSIGHT, a data-driven survival framework that learns spatially resolved risk from routine haematoxylin and eosin whole-slide images. Cross-validated on multicentre cohorts (n = 678; TCGA, SurGen) and independently validated in PETACC-3 (n = 1155) and Orion (n = 27), INSIGHT outperformed pTNM staging (C-index 0.68 vs 0.58; hazard ratio 4.00 [2.85–5.62] vs 2.54 [1.67–3.89]) and remained the strongest independent predictor after adjustment for clinical and previously known molecular signatures.
Using spatially localised risk as an outcome surrogate, we integrated predictions with single-cell spatial transcriptomics (n = 15), multiplex immunofluorescence (n = 27), immunohistochemistry  (n = 80), bulk transcriptomics (n = 1497), and single-cell reference data. This revealed an epithelial-immune risk manifold defined by epithelial dedifferentiation and fetal-like programmes, myeloid-driven immunosuppression, and adaptive immune dysfunction, consistent with mechanistic models.
A substantial fraction of molecular prognostic signal was recoverable from tissue architecture, with complementary clinicomolecular features improving performance to a C-index of $0.74 \pm 0.03$.
Together, INSIGHT shows histology encodes core drivers of prognosis into spatially organised tissue states, providing a unified framework for prediction, biological discovery, and clinical translation.}
\end{abstract}

\maketitle

\section{Introduction}
Around 75\% of the 1.9 million new annual colorectal cancer (CRC) diagnoses worldwide are non-metastatic (stage I-III)~\cite{xi2021global}. For these patients, surgical resection remains the standard of care. However, approximately 25\% of stage II-III cases harbour undetected micro-metastatic disease at diagnosis, leading to cancer-related death~\cite{murray2020subtypes}. Prognosis, typically assessed through pathological Tumour-Node-Metastasis (pTNM) staging after surgery, guides adjuvant chemotherapy decisions. \textcolor{black}{However, pTNM provides limited prognostic resolution and offers incomplete treatment guidance, particularly in stage II/III disease where uncertainty remains high~\cite{liu2018p}.} Efforts to refine risk stratification using various features have shown limited benefit~\cite{chen2021pathological}. Recent studies show that image-derived predictors can stratify risk from histopathology, but often depend on non-image clinical variables and offer limited biological interpretability~\cite{kleppe2022clinical,skrede2020deep}. Circulating tumour DNA (ctDNA) shows promise, but is not mature enough for clinical integration~\cite{ma2025circulating}. Altogether, current prognostication frameworks put thousands of patients at risk for both over- and undertreatment.

\textcolor{black}{Motivated by this unmet need in stage II/III CRC, we focus on two questions: (i) whether clinically actionable prognostic information can be derived from routine H\&E histology used in standard workflows, and (ii) what spatially organised tumour programmes underlie risk. We posit that routine histology encodes integrated prognostic information arising from the combined effects of tumour-intrinsic and microenvironmental processes. Conventional mechanistic approaches are not well-suited to capture this signal, as they interrogate predefined features in isolation. Conversely, cancer progression is shaped by spatially and temporally heterogeneous hallmark-associated processes spanning tumour cell states, microenvironmental interactions, and systemic influences~\cite{hanahan2026hallmarks}. Prognostic biology emerges from the coordinated interplay of epithelial state, stromal remodelling, immune regulation, and tissue architecture, giving rise to higher-order patterns not accessible to reductionist analyses. Although emerging multi-omic and spatial profiling technologies resolve this complexity at various scales, integrating high-dimensional data to prognosticate remains challenging. Capturing such structure requires data-driven approaches that integrate information across scales.}

\textcolor{black}{Under this premise, we introduce INSIGHT, a graph neural network (GNN) trained on large-scale whole-slide image (WSI) cohorts with survival outcomes. By operating directly on WSIs, INSIGHT is clinically applicable without additional assays. Each slide is modelled as a graph of spatially connected patches, enabling multi-scale, data-driven characterisation of tissue architecture. The learned representations capture prognostic signals that both recapitulate known correlates and reveal additional patterns associated with outcome. INSIGHT captures both local and higher-order spatial structure and produces complementary outputs: patient-level risk scores for stratification and localised patch-level risk maps that resolve tissue risk.}

\textcolor{black}{Crucially, INSIGHT's spatially localised risk can act as surrogate survival labels, enabling orthogonal modality integration. This allows systematic identification of gene-, pathway-, and protein-level outcome correlates, linking image-derived patterns to tumour biology. INSIGHT thus moves beyond conventional image-based prediction and reductionist, single-mechanism approaches by using large-scale survival supervision to recover biologically meaningful programmes from histology, providing a scalable, data-driven bridge between prognostic modelling and mechanistic interpretation.}

Our key contributions are:

\begin{enumerate}

\item INSIGHT, a GNN that prognosticates from routine H\&E WSIs by modelling tissue as unbiased spatial patch graphs, producing both patient-level and spatially resolved risk scores. This enables clinically deployable prognostic stratification in stage II/III CRC and provides localised surrogate survival signals for multimodal biological integration.

\item Large-scale multicentre training (TCGA, SurGen; $n=678$) with independent external validation on PETACC-3 ($n=1155$) and Orion ($n=27$) demonstrates superior prognostic performance relative to pTNM staging and contemporary histology-based models (C-index $0.68$-$0.69$ vs $0.44$-$0.58$). Multivariable Cox analysis shows that INSIGHT provides independent prognostic value beyond clinical variables and established molecular programmes (PETACC-3: HR $1.51[1.24-1.85]$).

\item Integrating INSIGHT spatial risk with single-cell spatial transcriptomics (IMMUcan, $n=15$; 422 genes), multiplex immunofluorescence (Orion, $n=27$), immunohistochemistry (Janssen, $n=80$), and bulk transcriptomics (TCGA, PETACC-3) identifies a coherent epithelial-immune risk manifold spanning epithelial dedifferentiation and fetal-like programmes, myeloid-driven stromal states ($\mathrm{SPP1}^{+}$ macrophages and $\mathrm{LAMP3}^{+}$ dendritic cells), and adaptive immune dysfunction. These axes are consistently observed across modalities and align with independent mechanistic studies, including organoid and in-vivo models of microenvironment-driven tumour progression, demonstrating concordance with independent biological evidence.

\item INSIGHT recapitulates canonical prognostic histopathological features while identifying novel risk correlates, including nuclear solidity and circularity linked to fetal-like epithelial states. It reveals heterogeneity, including context-dependent LGR5-associated epithelial states,  MSI-High stratification into low-risk stem-like and high-risk CD8 T-cell exhausted phenotypes, and distinct high-risk epithelial programmes associated with loss of CDX2/HNF4A and CEACAM5/6-driven pathways.

\end{enumerate}

\begin{figure}
    \centering
    \includegraphics[width=1\linewidth]{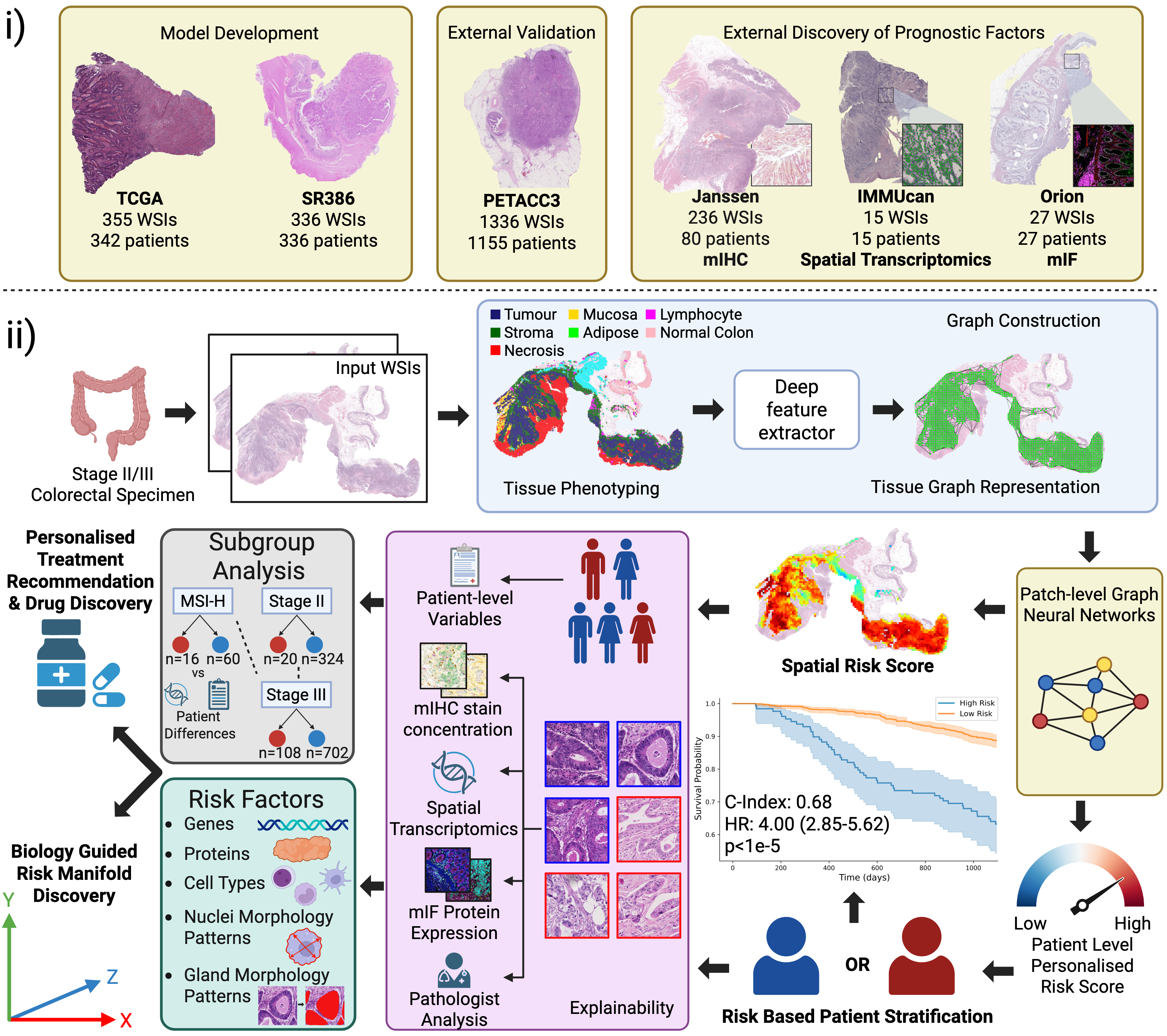}
    \caption{Overview of the overall INSIGHT pipeline. Part i) shows the summary of different H\&E WSI datasets used for model development and validation. Additional multi-modal datasets containing both H\&E and an additional WSI modality were used for downstream discovery of prognostic factors associated with patient prognosis. The overview of the pipeline which utilizes H\&E WSIs from Stage II/III Colorectal (CRC) specimens can be seen in ii).  For each WSI we first perform tissue phenotyping to automatically restrict our analysis to tumour, stroma, and lymphocyte regions.  These regions are then broken down into square patches for which we extract deep features. These are then used to construct a tissue graph, based on patch proximity. Such graphs can be used to train a Graph Neural Network (GNN) that directly predicts a patient-level personalized risk score. These risk scores can rank and stratify patients, as shown for the external test set PETACC-3 via the Kaplan-Meier Curve. Further, we can deconvolve the patient risk score into spatial patch-level risk. This can be seen in the WSI heatmap (red and blue signifying high and low risk respectively).  These patch-level risk scores can be analysed by a pathologist as well as act as an automatic survival label. This allows us to discover patterns within different modalities that may not have a ground truth survival label. Here we analyse the relationship between spatial risk and various factors derived from mIHC, mIF and Spatial Transcriptomics from three external cohorts. This allows us to discover, in a time efficient manner, what biologically interpretable factors are most associated with prognosis. Further, patient-level risk scores allow performing subset discovery within clinical populations. These discoveries can be used to guide treatment decisions and future drug development that target these high-risk features. Further, we can combine all our risk factors into a single risk manifold that is biologically grounded and explains away INSIGHTS risk predictions}
    \label{fig:main_figure}
\end{figure}

\section{Results}
\subsection{\textcolor{black}{INSIGHT achieves robust prognostic performance across multicentric stage II/III CRC cohorts}}

We evaluated INSIGHT’s prognostication ability (see Fig ~\ref{fig:main_figure}) across diverse clinical settings. Four-fold cross-validation was performed on 691 WSIs from 678 patients, spanning multiple cohorts, scanners, and institutions (Supplementary Table 1), with the training cohort comprising TCGA (n = 342) and SurGen (n = 336). INSIGHT achieved a Concordance Index (C-index) of 0.67 and a hazard ratio (HR) of 2.62 (95\% CI: 1.80–3.81), compared to 0.60 and 2.31 (95\% CI: 1.54–3.47) for pTNM staging (Table~\ref{tab:main_results}).

Subpopulation analyses,assessing robustness, identified performance was highest in MSI-High, female, and younger patients, while stage II cases showed lower discrimination (C-index = 0.62; HR = 2.26, 95\% CI: 1.14–4.47).

We evaluated generalisability on independent cohorts. In PETACC-3 ($n = 1155$), INSIGHT achieved a C-index of 0.68 and HR of 4.00 (95\% CI: 2.85–5.62) (Table~\ref{tab:main_results}). Kaplan–Meier analysis confirmed significant stratification between risk groups ( Fig.~\ref{fig:TCGA_km}; Extended Fig.~\ref{fig:petacc_km}). Performance exceeded that of pTNM staging (C-index 0.58; HR 2.54, 95\% CI: 1.67–3.89) and multiple-instance learning approaches (C-index 0.64; HR 2.55, 95\% CI: 1.82–3.58). This was consistent across evaluated subpopulations. In Orion ($n=27$), INSIGHT achieved a C-index of 0.69 and HR of 2.23 (95\% CI: 0.43–11.48), compared to 0.45 and 0.66 (95\% CI: 0.15–2.94) for pTNM staging.

\textcolor{black}{Importantly, INSIGHT retained independent prognostic value beyond clinical and established molecular variables in multivariable analysis (Table~\ref{tab:main_results} and Section~\ref{sec:multivariable}). Results summary can be found at: \href{https://insight.dcs.warwick.ac.uk/}{https://insight.dcs.warwick.ac.uk/}}. 

\begin{table}[]
\centering
\begin{threeparttable}
\resizebox{0.9\textwidth}{!}{
\begin{tabular}{llccccc ccccccc|cc}\toprule
        
\textbf{Variable} & \textbf{Group}
& \multicolumn{2}{c}{\textbf{Stage Risk (Baseline)}}
& && \multicolumn{2}{c}{\textbf{DeepAttnMISL Risk~\cite{yao2019deep}}}
& && \multicolumn{2}{c}{\textbf{MHAttnSurv Risk~\cite{jiang2023mhattnsurv}}}
& && \multicolumn{2}{c}{\textbf{INSIGHT Risk (Proposed)}}\\

\cline{3-4}\cline{7-8}\cline{11-12}\cline{15-16}

& & C-index $\pm$ SD& HR (95\% CI) p-value
& && C-index $\pm$ SD& HR (95\% CI) p-value
& && C-index $\pm$ SD& HR (95\% CI) p-value
& && C-index $\pm$ SD& HR (95\% CI) p-value\\
\midrule

\multicolumn{16}{c}{\textbf{Internal Test Set (TCGA+SR386)}}\\
\multicolumn{16}{c}{\textbf{691 WSIs from 678 Patients}}\\
\midrule

Entire Cohort& 
& $0.60 \pm 0.02$& $2.31 (1.54 - 3.47)$ $p<0.0001$
& & & $\textbf{0.67} \pm \textbf{0.02}$& $2.74 (1.87-4.0)$ $p<0.0001$
& & & $0.66 \pm 0.03$& $2.89 (1.99-4.19)$ $p<0.0001$
& & & $\textbf{0.67} \pm \textbf{0.03}$& $2.62 (1.80-3.81)$ $p<0.0001$
\\

Stage & & & & & & & & & & & & & & &\\

& Stage II $(n=332)$
& -&-
& & & $\textbf{0.68} \pm \textbf{0.05}$&$3.22 (1.61-6.43)$ $p=0.0004$
& & & $0.67 \pm  0.05$&$4.04 (2.05-7.94)$ $p<0.0001$
& & & $0.62 \pm 0.05 $&$2.26 (1.14-4.47)$ $p = 0.016$
\\

& Stage III $(n=346)$
& -&-
& & & $0.66 \pm 0.03$&$2.31 (1.46-3.64)$ $p=0.0002$
& & & $0.65 \pm 0.03$&$2.27 (1.45-3.55)$ $p=0.0002$
& & & $\textbf{0.68} \pm \textbf{0.03}$&$2.72 (1.74-4.26)$ $p<0.0001$
\\

MSI Status& & & & & & & & & & & & & & &\\

& MSS $(n=593)$
& $0.60 \pm 0.02$&$2.39 (1.55-3.69)$ $p<0.0001$
& & & $\textbf{0.67} \pm \textbf{0.03}$&$2.77 (1.86-4.13)$ $p<0.0001$
& & & $0.66 \pm 0.03$&$2.95 (2.0, 4.35)$ $p<0.0001$
& & & $\textbf{0.67} \pm \textbf{0.03}$&$2.64 (1.79-3.91)$ $p<0.0001$
\\

& MSI High $(n=79)$
& $0.55 \pm 0.09$&$1.46 (0.37 - 5.86)$
& & & $\textbf{0.80} \pm \textbf{0.06}$&$3.46 (0.93-12.9)$ $p=0.05$
& & & $0.72 \pm 0.09$&$3.18 (0.79-12.71)$
& & & $0.73 \pm 0.09$&$3.67 (0.76-17.69)$
\\

Sex& & & & & & & & & & & & & & &\\

& Female $(n=332)$
& $0.61 \pm 0.03$&$2.80 (1.49-5.24)$ $p=0.0008$
& & & $0.67 \pm 0.04$&$2.42 (1.38-4.23)$ $p=0.001$
& & & $\textbf{0.69} \pm 0.04$&$3.59 (2.08-6.2)$ $p<0.0001$
& & & $0.68 \pm 0.04$& $2.83 (1.64-4.88)$ $p<0.0001$
\\

& Male $(n=355)$
& $0.59 \pm 0.03$&$2.00 (1.17-3.40)$ $p=0.009$
& & & $\textbf{0.68} \pm \textbf{0.04}$&$3.11 (1.85-5.2)$ $p<0.0001$
& & & $0.64 \pm 0.04$&$2.34 (1.4-3.9)$  $p=0.0008$
& & & $0.66 \pm 0.04$& $2.46 (1.47-4.12)$ $p=0.0004$
\\

Age& & & & & & & & & & & & & & &\\

& $>60$ $(n=475)$
& $0.60 \pm 0.03$&$2.23 (1.45-3.44)$ $p=0.0001$
& & & $\textbf{0.65} \pm \textbf{0.03}$&$2.24 (1.46-3.42)$ $p=0.0001$
& & & $0.63 \pm 0.03$&$2.57 (1.7-3.91)$ $p<0.0001$
& & & $0.63 \pm 0.03$& $2.12 (1.40-3.22)$ $p=0.0003$
\\

& $\leq60$ $(n=203)$
& $0.64 \pm 0.04$&$6.24 (1.46 - 26.69)$ $p=0.005$
& & & $0.74 \pm 0.06$&$5.08 (2.16-11.91)$ $p<0.0001$
& & & $0.74 \pm 0.05$&$3.67 (1.58-8.50)$ $p=0.001$
& & & $\textbf{0.79} \pm \textbf{0.05}$& $4.92 (2.12-11.39)$ $p<0.0001$
\\

\midrule

\multicolumn{16}{c}{\textbf{External Test Set (PETACC-3)}}\\
\multicolumn{16}{c}{\textbf{1336 WSIs from 1155 Patients}}\\
\midrule

Entire Cohort& &
$0.58 \pm 0.02$&$2.54 (1.67-3.89)$ $p<0.0001$
& && $0.64 \pm 0.02$&$2.55 (1.82-3.58)$ $p<0.0001$
& && $0.62 \pm 0.02$&$2.66 (1.76-4.03)$ $p<0.0001$
& && $\textbf{0.68} \pm \textbf{0.02}$&$4.00 (2.85-5.62)$ $p<0.0001$
\\

Stage & & & & & & & & & & & & & & &\\

& Stage II $(n=345)$
& -&-
& && $0.53 \pm 0.06$& $1.03 (0.31-3.46)$
& && $\textbf{0.61} \pm \textbf{0.05}$&$1.58 (0.37-6.70)$
& && $0.60 \pm 0.06$&$2.48 (0.74-8.33)$
\\

& Stage III $(n=810)$
& -&-
& && $0.66 \pm 0.02$& $2.75 (1.92-3.94)$ $p<0.0001$
& && $0.61 \pm 0.02$&$2.66 (1.73-4.11)$ $p<0.0001$
& && $\textbf{0.68} \pm  0.02$&$3.80 (2.65-5.44)$ $p<0.0001$
\\

MSI Status& & & & & & & & & & & & & & &\\

& MSS $(n=939)$
& $0.57 \pm 0.02$&$2.26 (1.43-3.57)$ $p=0.0004$
& && $0.63 \pm 0.03$& $2.34 (1.55-3.53)$ $p<0.0001$
& && $0.61 \pm 0.03$&$2.62 (1.61-4.27)$ $p<0.0001$
& && $\textbf{0.67} \pm \textbf{0.02}$&$3.36 (2.21-5.12)$ $p<0.0001$
\\

& MSI High $(n=140)$
& -&-
& && $0.678\pm 0.10$& $ 3.04 (0.68-13.59)$
& && $0.71 \pm 0.10 $&$2.61 (0.51-13.44)$
& && $\textbf{0.76} \pm 0.09$&$5.50 (1.23-24.62)$ $p=0.012$
\\

Sex& & & & & & & & & & & & & & &\\

& Female $(n=512)$
& $0.58 \pm 0.02$&$2.36 (1.20-4.63)$ $p=0.01$
& && $0.69 \pm 0.03$ & $3.02 (1.8-5.07)$ $p<0.0001$
& && $0.66 \pm 0.04 $&$2.85 (1.48-5.45)$ $p=0.0009$
& && $\textbf{0.73} \pm \textbf{0.03}$& $6.03 (3.62-10.03)$ $p<0.0001$
\\

& Male $(n=643)$
& $0.58 \pm 0.02$&$2.69 (1.58-4.65)$ $p=0.0002$
& && $0.61 \pm 0.03$ & $ 2.3 (1.46-3.62)$ $p=0.0002$
& && $0.59 \pm 0.03$&$ 2.54 (1.49-4.34)$ $p=0.0004$
& && $\textbf{0.64} \pm \textbf{0.03}$& $2.95 (1.85-4.71)$ $p<0.0001$
\\

Site& & & & & & & & & & & & & & &\\

& Left $(n=767)$
& $0.57 \pm 0.02$&$2.49 (1.39-4.48)$ $p=0.002$
& && $0.61 \pm 0.03$& $ 1.86 (1.09-3.15)$ $p=0.02$
& && $0.61 \pm 0.04$&$ 2.42 (1.29-4.56)$ $p=0.005$
& && $\textbf{0.66} \pm \textbf{0.03}$&  $3.14 (1.83-5.41)$ $p<0.0001$
\\

& Right $(n=388)$
& $0.60 \pm 0.02$&$2.77 (1.50-5.13)$ $p=0.0007$
& && $0.66 \pm 0.03$ & $ 2.92 (1.84-4.63)$ $p<0.0001$
& && $0.61 \pm 0.03$&$2.5 (1.44-4.34)$ $p=0.0008$
& && $\textbf{0.68} \pm \textbf{0.03}$& $4.07 (2.57-6.45)$ $p<0.0001$
\\

Age& & & & & & & & & & & & & & &\\

& $>60$ $(n=572)$
& $0.58 \pm 0.02$&$2.50 (1.46-4.27)$ $p=0.0005$
& && $0.62 \pm 0.03$ & $ 1.99 (1.24-3.21)$ $p=0.03$
& && $0.58 \pm 0.03$&$1.91 (1.04-3.49)$ $p=0.03$
& && $\textbf{0.65} \pm \textbf{0.03}$& $2.73 (1.66-4.46)$ $p<0.0001$
\\

& $\leq60$ $(n=583)$
& $0.58 \pm 0.02$&$2.75 (1.37-5.54)$ $p=0.003$
& && $0.68 \pm 0.03$ & $3.51 (2.14-5.75)$ $p<0.0001$
& && $0.67 \pm 0.04$&$ 3.87 (2.18-6.87)$ $p<0.0001$
& && $\textbf{0.73} \pm \textbf{0.03}$& $6.28 (3.85-10.23)$ $p<0.0000$
\\

\midrule
\multicolumn{16}{c}{\textbf{External Test Set (Orion)}}\\
\multicolumn{16}{c}{\textbf{27 WSIs from 27 Patients}}\\
\midrule

Entire Cohort&
& $0.45 \pm 0.11$& $0.66 (0.15-2.94)$
& && $0.67 \pm 0.11$& $2.67 (0.52-13.76)$
& && $\textbf{0.83} \pm 0.08$& $3.6 (0.7-18.6)$
& && $0.69 \pm 0.11$&$2.23 (0.43-11.48)$
\\

\midrule
\multicolumn{16}{c}{\textbf{Multivariable Cox Models with Clinical + Molecular Covariates and Risk Predictors (PETACC-3)}}\\
\multicolumn{16}{c}{\textbf{543 WSIs from 543 Patients}}\\
\midrule
Entire Cohort&
& $0.70 \pm 0.02$& $0.84 (0.70-1.03)$
& && $0.72 \pm 0.04$& $1.39 (1.25-1.55)$ $p<0.005$
& && $0.70 \pm 0.05$& $ 1.27 (1.25-1.29)$ $p<0.005$
& && $\textbf{0.74} \pm \textbf{0.03}$&$1.51 (1.47-1.56)^{\dagger}$ $p<0.005$
\\

\bottomrule 
\end{tabular}}
\begin{tablenotes}
\footnotesize
\item[$\dagger$] This feature exhibited the highest hazard ratio in the multivariable Cox analysis.
\end{tablenotes}
\end{threeparttable}
\caption{Comparison of the INSIGHT model with pTNM stage–based stratification and contemporary multiple instance learning (MIL) approaches for survival prediction (overall survival censored at 3 years). Results are reported on an internal test set using 4-fold cross-validation (TCGA and SR386) and on external validation cohorts (PETACC-3 and Orion). Performance metrics include the concordance index (C-index, mean ± SD from 1000 bootstraps) and hazard ratio (HR) with 95\% confidence intervals and log-rank p-values. Results are shown for the overall cohort and clinically relevant subgroups (stage, MSI status, sex, age, andtumourr site). Across most settings, INSIGHT achieves competitive or superior performance compared to pTNM staging and MIL-based methods, indicating improved risk stratification and generalizability. For the PETACC-3 cohort, multivariable Cox proportional hazards models are constructed using clinical and molecular covariates together with each model-derived risk score. Risk predictors are generated by models trained exclusively on TCGA and SR386 and are therefore not fitted on PETACC-3. To mitigate overfitting in the Cox models, we perform 5-fold cross-validation on PETACC-3 and report the mean C-index alongside hazard ratios with 95\% confidence intervals and log-rank p-values. In this setting, INSIGHT achieves the strongest performance, suggesting that it captures prognostic information complementary to established clinicopathological and molecular features.}
\label{tab:main_results}
\end{table}

\subsection{\textcolor{black}{Spatial risk maps recover histopathological correlates of prognosis}}
\begin{figure}
    \centering
    \includegraphics[width=0.7\linewidth]{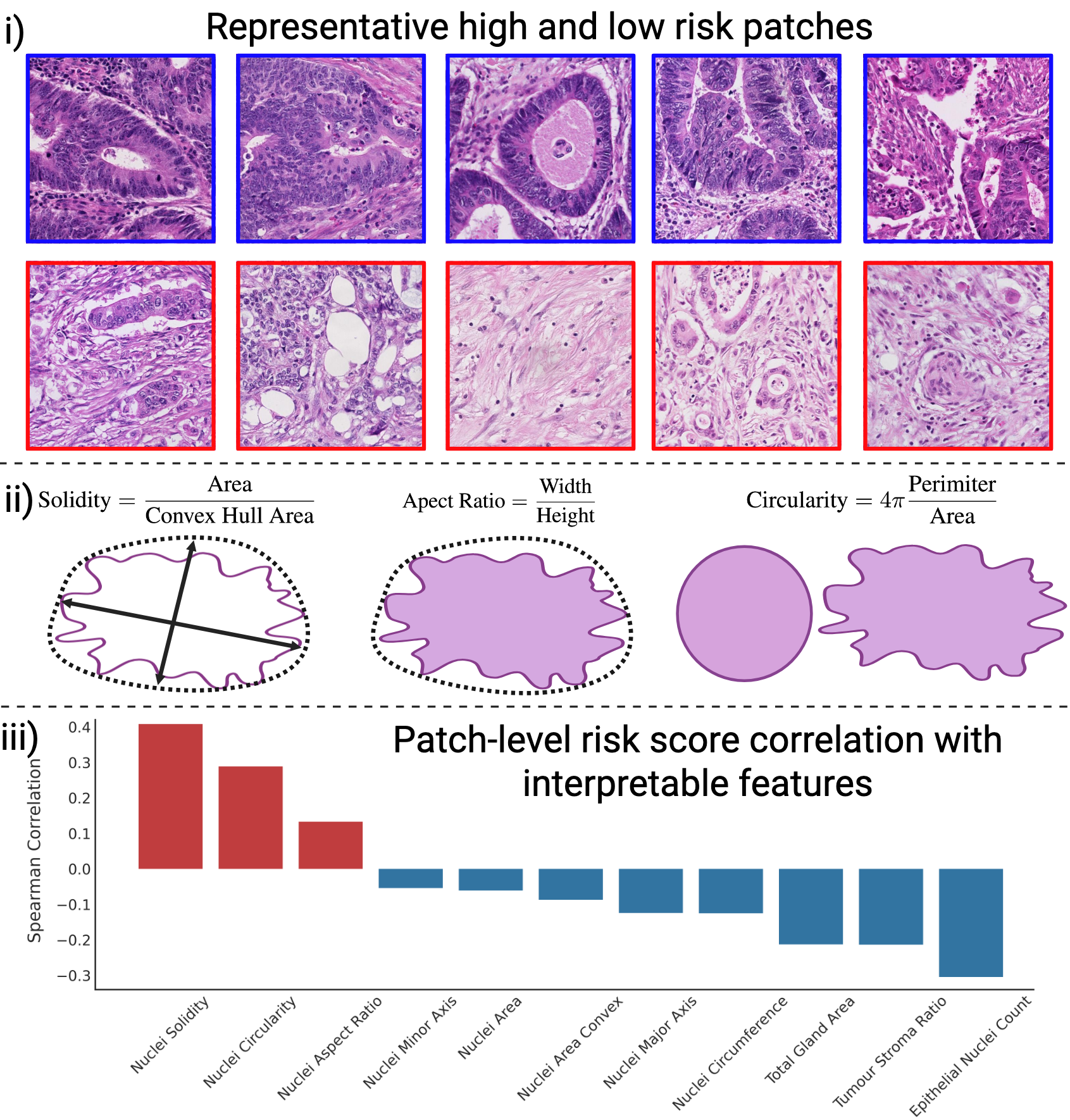}
    \caption{Morphological pattern analysis in high and low risk regions in external cohorts. In i) we see the representative high (red) and low (blue) risk patches taken from PETACC-3 where we see high risk regions showing a greater proportion of stroma and less well differentiated glands. In ii) we define the most important metrics in this study including nuclei solidity, aspect ratio and circularity that capture nuclei atypica. In iii) we systematically measure the relationship between patch-level risk and interpretable cell and gland features automatically generated using Cerberus in the Janssen cohort. This can be visualized via the bar chart showing the overall spearman correlation between a feature and localized risk.}
    \label{fig:histomorph_correlates}
\end{figure}

\textcolor{black}{Before using INSIGHT for pathobiological discovery, we first assessed its alignment with tumour biology, providing an important form of biological validation.} For this, we generated patch-level risk attributions and corresponding WSI heatmaps highlighting high- and low-risk regions. A consultant pathologist reviewed representative PETACC-3 patches (Fig.~\ref{fig:histomorph_correlates}i and Extended Fig.~\ref{fig:high_low_patches}), together with three randomly selected WSIs spanning low-, intermediate-, and high-risk patients. This qualitative assessment confirmed low-risk regions were characterised by well-formed glands, higher epithelium-to-stroma ratios, whereas high-risk regions showed increased stromal content, poorer differentiation, and nuclear atypia. 

Complementing qualitative analysis, we systematically quantified the relationship between patch-level risk and interpretable morphometric features derived from automated cell and gland segmentations~\cite{graham2023one}. Fig.~\ref{fig:histomorph_correlates}ii defines the key nuclear shape descriptors analysed, including solidity, aspect ratio, and circularity, which capture nuclear irregularity and pleomorphism. Higher-risk patches, Fig.~\ref{fig:histomorph_correlates}iii, exhibited strong correlations with increased nuclear solidity ($\rho=0.41$) and circularity ($\rho = 0.29$), indicative of more compact morphology, consistent with nuclear deformation and poor prognosis links~\cite{bertillot2025colorectal}, whereas lower-risk patches showed increased tumour to stroma ratio ($\rho=-0.21  \text{ to} -0.30$), see Extended Fig.~\ref{fig:janssen_vis}.

\subsection{\textcolor{black}{Spatial transcriptomics links localised risk to epithelial identity loss, fetal-like programmes, and context-dependence}}
Building on histomorphological analysis, we used INSIGHT-derived patch-level risk maps to identify spatial molecular correlates. This enabled molecular discovery in cohorts with spatial profiling but limited survival information. We applied this approach to the IMMUcan cohort ($n=15$), which had 422 gene cell-level spatial transcriptomic profiles with a co-registered serial H\&E on which INSIGHT risk scores were computed, enabling identification of gene-level prognostic correlates.

\begin{figure}
    \centering
    \includegraphics[width=0.7\linewidth]{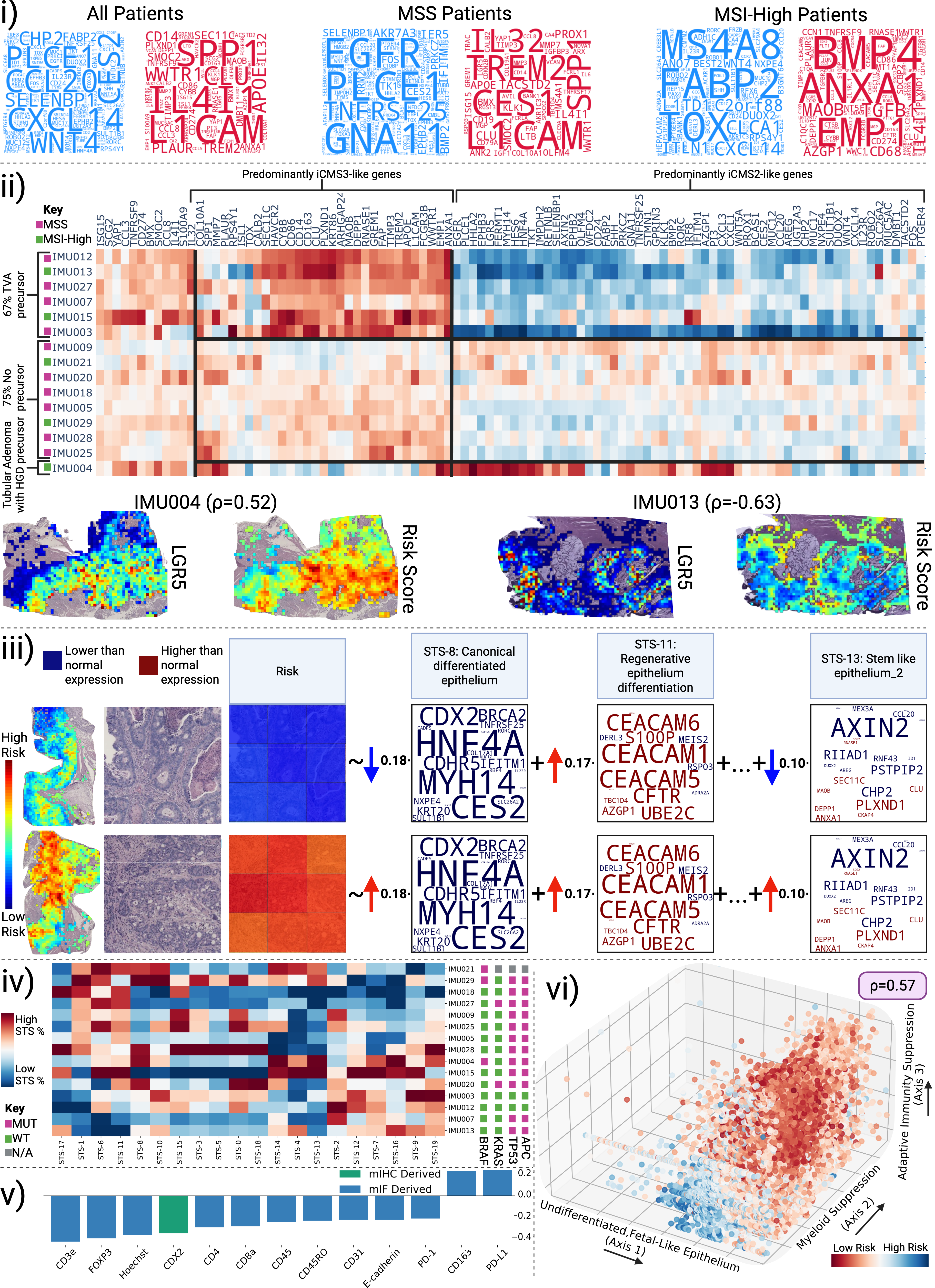}
    \caption{Association of spatially localized risk with multi-modal patch-level factors. In i)-iv) we focus on spatial transcriptomic profiles. In i) we begin with univariate analysis between patch-level gene expression derived from spatial transcriptomics and INSIGHT generated patch-level risk scores per patient. This is shown via the heatmap that shows the overall correlation between patient patches. To analyze finer grained detailed at a patient-level  we show heatmap of patch-level risk scores and scaled patch-level gene expression within a patient, ii). This allows us to validate nuanced patient-level patterns, for example the opposite correlation of LGR5 to prognosis in IMU013 and IMU004. In iii) we perform multivariate gene analysis to regress risk with respect to data-driven patch-level Spatial Transcriptomic Signatures (STS) derived from 422 spatial transcriptomic genes. A given signature is represented via a word cloud, where red and blue genes represent if it is over- or underexpressed in a signature and the size of the gene represents its importance to the signature definition.  We can then regress the binary signatures against localised risk via linear regression (OLS) allowing us to find the most important STSs, in this case STS- 8, STS -11 and STS -13. In iv) to further understand the STS expression patterns we show the proportion of the total patches in a patient that are activated for each signature, depicted via a heatmap, with blue and red depicting small and large proportions, respectively. For each patient, we also show several key point mutations and whether they are mutated (purple) or not (green).  In v) we show univariate analysis between localised predicted risk and concentration of mIHC stains and mIF proteins as additional validation of our discoveries. This is visualised as a bar chart depicting the strength of correlations for different proteins. Finally, in vi) we show our final derived risk manifold based on combining STS signatures, mIF proteins and micromorphological factors. With dimensionality reduction (see Methods), we can combine all these risk-associated features into a single risk manifold with three main axes. This is shown on the scatter plot, where each point represents a single patch, coloured by its INSIGHT risk.}
    \label{fig:spatial_analysis}
\end{figure}

Cohort level uinivariate analysis (Fig.~\ref{fig:spatial_analysis}i), showed high-risk regions significantly correlated with increased expression of injury repair genes L1CAM\cite{moorman_progressive_2025} ($\rho=0.29$) and IL4I1 ($\rho=0.28$), and protumorigenic macrophage marker SPP1 ($\rho=0.26$), and negatively correlated with immune-activating chemoattractants including CXCL14 ($\rho=-0.41$), CCL20 ($\rho=-0.39$) and absorptive lineage marker CES2 ($\rho=-0.37$). While these aggregate trends highlight recurrent prognostic programs, they can mask inter-patient variability. Patient-level analyses (Fig.~\ref{fig:spatial_analysis}ii) revealed genes with heterogeneous risk associations. Notably, LGR5, a canonical marker for Wnt-high colorectal cancer stem cells\cite{hong_epithelial_2025-1,white2026mapk,mzoughi_oncofetal_2025}, exhibited inverse correlations, with a negative and positive association in IMU013 ($\rho=-0.63$) and IMU004 ($\rho=0.52$), respectively. Corresponding spatial heatmaps confirmed this divergence. This highlights the importance of patient-level spatial analysis for interpreting prognostic signals, which may be obscured by cohort-averaged analyses. 

Beyond gene-risk associations, we sought to identify risk-associated coordinated spatial gene expression programs. Using Correlation Explanation~\cite{pepke2016multivariate}, a data-driven approach that captures multivariate gene-expression dependencies without predefined pathway reliance, we identified 20 recurring spatial transcriptomic patterns, which we term Spatial Transcriptomic Signatures (STS) (Fig.~\ref{fig:spatial_analysis}iii, Extended Fig.~\ref{fig:topics_part_1} and~\ref{fig:topics_part_2} and Extended Table~\ref{tab:ols_results}). Each STS represents a weighted combination of genes and captures a coordinated mode of gene expression that assumes a binary state at each spatial location, corresponding to two opposing expression configurations. We visualised each STS using word clouds where gene size reflects its contribution weight and colour indicates relative expression (red = higher, blue = lower) when the STS is active. For example, activation of STS-11 corresponds to CEACAM6/5 overexpression and MEIS2/RSPO3 underexpression, with the inverse when inactive.
\textcolor{black}{STS-8 emerged as the dominant contributor to localised prognostic variation (18\%) as assessed by regressing localised risk against binary STS activation states. STS-8, which downregulated genes from known LGR5 ISC signatures, absorptive intestine ~\cite{moorman2025progressive}, and iCMS2 and CHRIS-C subtypes\cite{joanito_single-cell_2022}~\cite{isella2017selective} was characterized by  a shift toward fetal like states. STS-11 represented the second largest contributor (17\%), upregulated iCMS2, CHRIS-C-specific~\cite{isella2017selective}, colonocyte, intestinal and few high relapse genes~\cite{moorman2025progressive}, and was defined by CEACAM5/6 overexpression, indicating a distinct regenerative epithelium differentiation program. CDX2 marks colonic lineage identity and its downregulation is a key feature of STS-8 activation, correlating with risk. Alternatively, CDX2 expression was moderately upregulated in STS-11, indicating alternative pathways and modes to achieve disease progression with or without epithelial lineage identity, supported by CDX2 oncogenic and tumour suppressor functions~\cite{salari2012cdx2}. Moreover, as these spatial transcriptomics signatures are derived at the patch-level, INSIGHT can also inform non-epithelial genes spatially coregulated with epithelial programs; STS-8 co-downregulates RORC, a TH17 differentiation regulator, and IL23R, a pro-inflammatory gene found on immune cells, indicating non-epithelial cells expressing these genes may support the loss of colonic identity in tumours. STS-11, co-downregulated B cell (Cd79A/B, NS4A1, BANK1, FCRL1), T cell (TRAT1, ITK), exhaustion (CTLA4, EOMES), endothelial (LYVE1, ACKR1, AQP1) and ECM regulatory (FAP, ACTA2, COL10A1, VCAN, THBS1, TIMP3)-specific genes, while upregulating T regulatory cell marker FOXP3. Since CTLA4 and LYVE1 is generally associated with worse CRC prognosis~\cite{derakhshani2021cytotoxic}, their co-downregulation in STS-11 indicates differential tumour identity and non-epithelial programs associated with risk that may not be only patient-, but tumour location-specific. WSI analysis showed variability in the proportion of patches with activation of different STS’s, (Fig.~\ref{fig:spatial_analysis}iv). Notably, all patients with a high proportion of STS-8 activated patches were BRAF-mutated or BRAF-like, suggesting an association between STS and underlying genomic alterations. }

\subsection{\textcolor{black}{Spatial protein profiling validates epithelial, myeloid, and T-cell correlates of risk}}
To identify spatial protein programs associated with histology-derived risk, we analysed Janssen ($n = 80$) and Orion ($n = 27$) cohorts with serial tissue sections where INSIGHT risk predictions from H\&E slides could be spatially aligned with protein measurements from registered mIHC and mIF sections. This revealed consistent risk–protein associations (Fig.~\ref{fig:spatial_analysis}v). High-risk regions exhibited reduced expression of CDX2 ($\rho=-0.34$), T cell markers CD3e ($\rho=-0.42$), CD4 ($\rho=-0.29$) and CD8a ($\rho=-0.28$) alongside increased expression of macrophage marker CD163 ($\rho=0.23$) and T cell suppressor PD-L1 ($\rho=0.24$). While CD3E downregulation was the biggest contributor to STS-1, which encompasses a co-downregulation of cytotoxic T cell associated genes, CD3E was conversely upregulated in STS’s encompassing stromal (STS-2) and myeloid (STS-7) activation, suppressive immunity (STS-16), and exhausted niches (STS-9/12). In line with protein expression, CDX2 RNA downregulation was a major contributor to the highest risk-associated spatial program, STS-8. Together, these findings delineate spatial proteomic states linked to adverse prognosis, highlighting coordinated epithelial dedifferentiation and immune suppression at high-risk tissue sites.

\subsection{\textcolor{black}{A multimodal epithelial–immune risk manifold integrates spatial correlates of prognosis}}
To characterise the biological processes underlying INSIGHT, we integrated spatially resolved risk maps with matched molecular measurements across modalities, including spatial transcriptomics, multiplexed imaging, and bulk transcriptomic data. Across these, risk-associated features consistently organised into a low-dimensional structure defined by three recurrent biological themes: epithelial dedifferentiation with fetal-like states, myeloid-driven immunosuppression, and adaptive immune dysfunction.

To formalise this, we modelled localised risk as a projection onto these biologically defined axes using a multimodal risk manifold framework that aggregates features across modalities while accommodating missing data. This resolved local prognostic variation into three interpretable and non-overlapping axes (Fig.~\ref{fig:spatial_analysis}vi; Extended Fig.~\ref{fig:manifold_marginals} and Fig.~\ref{fig:manifold_weights}).

Axis 1, capturing epithelial dedifferentiation and fetal-like identity, showed strongest risk association ($\beta = 0.092$). Higher axis values were driven by fetal-like and regenerative epithelial programs, including STS-8 (canonical differentiated epithelium), STS-11 (regenerative epithelium differentiation), and STS-13 (stem-like epithelium), together with increased image-derived nuclear solidity. Axis 2, reflecting myeloid-driven immunosuppression, exhibited a smaller risk contribution ($\beta = 0.015$), characterised by increased suppressive immune activity, including STS-16 and elevated CD163 expression, indicating an additive role for myeloids in shaping prognosis. Axis 3, representing adaptive immune suppression, showed a substantial positive association with risk ($\beta = 0.076$) and was marked by reduced CD3e and CD8a expression alongside increased activity of cytotoxic T cell-downregulating signatures such as STS-1 and STS-12, consistent with dysfunctional immune states.

Together, these axes define a multimodal risk manifold that integrates epithelial and stromal contributions to prognosis, providing an interpretable framework for linking spatial tissue states to localised risk.

\subsection{INSIGHT reveals molecular and immune heterogeneity within MSI-high tumours}
\begin{figure}
    \centering
    \includegraphics[width=0.7\linewidth]{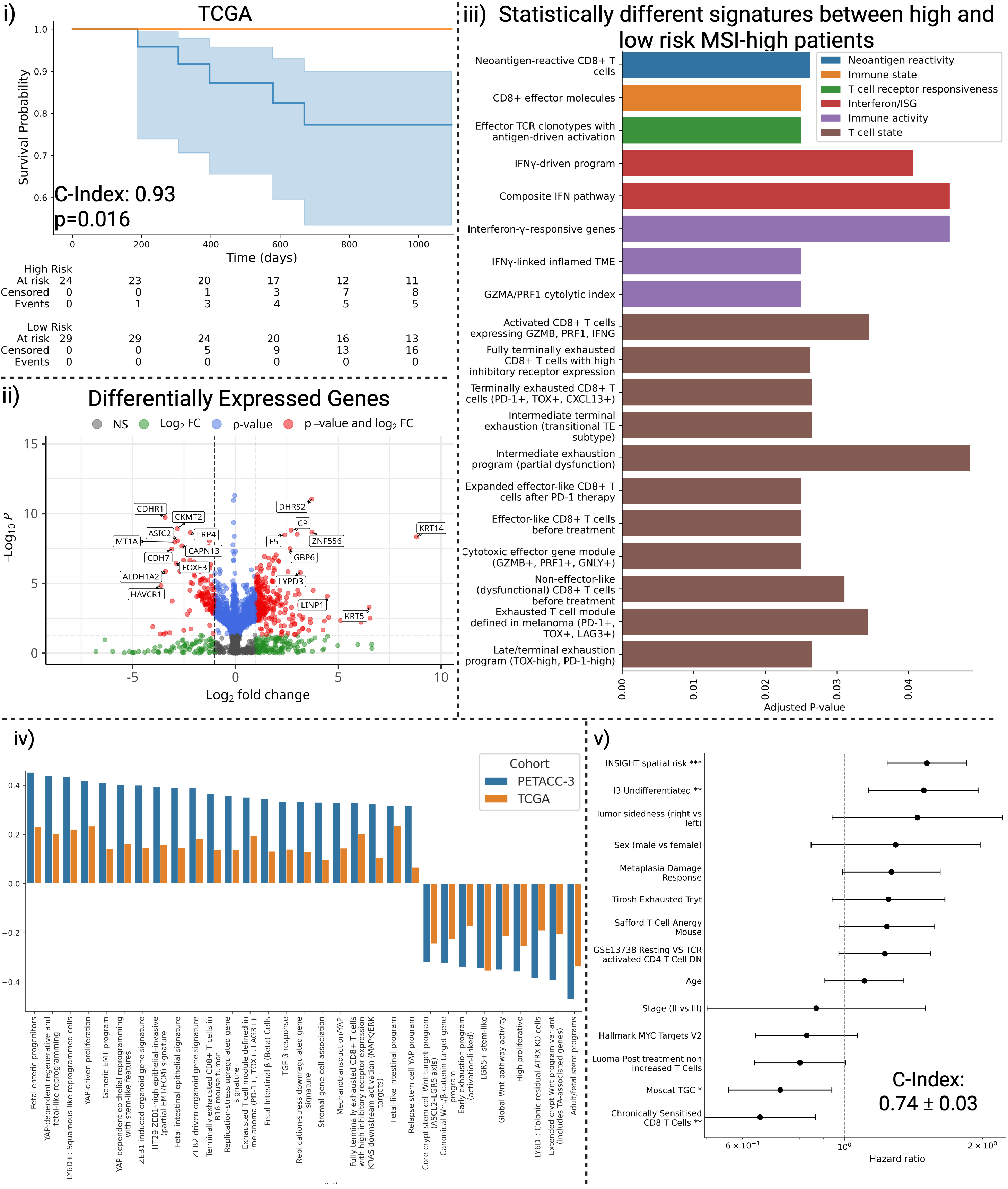}
    \caption{Patient level risk score analysis. In i)-iii) we explore the heterogeneity of MSI-High patients that was identified in Table 1. In i) we see in TCGA a strong patient ordering and separation in the Kaplan-Meir curve where MSI-High patients show a clear separation into two groups. The groups are defined based on the optimal threshold derived from the training set (see Methods). We then try to understand the groups difference via identifying differentially expressed genes, ii), as well as statistically significantly different bulk-derived signatures in iii) via a Mann-Whitney U test and coloured based on the pathway's main theme. In iv) we analyze the association between patient level pathways and risk scores across the whole Stage II/III CRC population in PETACC-3 and TCGA. We show the strongest correlations between Single Sample Gene Set Enrichment Analysis enrichment scores of various pathways and patient-level risk via a bar chart. v) Multivariable Cox model in PETACC-3. Forest plot showing hazard ratios (95\% CI) for INSIGHT risk, clinical variables, and bulk RNA-Seq derived pathway scores. INSIGHT spatial risk is the strongest independent prognostic factor, exceeding standard clinical variables. Residual significance of epithelial and immune pathways indicates complementary biological processes not fully captured by histology-derived risk.}
    \label{fig:msi_hetero}
\end{figure}

In Table~\ref{tab:main_results} and Fig.~\ref{fig:msi_hetero}i, we observe higher-than-average performance of INSIGHT in MSI-High patients (In TCGA: C-index = 0.93). This is notable, as MSI-High CRC is typically considered a relatively homogeneous and prognostically favourable subtype with important implications for treatment stratification~\cite{wang2019microsatellite}. This raises the possibility that meaningful, previously underappreciated heterogeneity exists here. This stratification was supported by marked transcriptional differences, with 527 differentially expressed genes (Fig.~\ref{fig:msi_hetero}ii). High-risk patients showed increased expression of keratin genes KRT14, previously linked to metastatic MSI-High CRC~\cite{xu2022analysis} KRT5, and CSAG1, whereas CEL, CDHR1, and Wnt-regulator LRP4 were underexpressed. 

Corroborating this, bulk RNA-seq-derived immune markers also differed significantly between MSI-High groups (Fig.~\ref{fig:msi_hetero}iii) including CD8-associated signatures such as the GZMA/PRF1 cytolytic index and markers of terminally exhausted $\mathrm{CD8}^{+}$ T cells ($\mathrm{PD-1}^{+}$, $\mathrm{TOX}^{+}$, $\mathrm{CXCL13}^{+}$)~\cite{thommen2018transcriptionally,hu2021cytolytic}. As contradictory T cell signaling states seem to be present in high-risk MSI-H patients concurrently, both the exhausted and activated score could reflect the number of tumor-infiltrating T cells, a low epithelium-to-immune ratio, or T cell dysfunction in high-risk MSI-High patients. These differences may occur between patients or even within individual tumors and could be associated with variation in patient risk.

\subsection{\textcolor{black}{Patient-level transcriptomics reveals fetal-like, lineage-loss, and immune-dysfunctional programmes underlying INSIGHT risk}}
To identify global transcriptional programs associated with patient-level risk, we analysed bulk RNA sequencing data from PETACC-3 and TCGA. Using curated gene signatures representing epithelial differentiation, fetal-like programs and immune states, we computed single-sample enrichment scores and correlated these with risk (Fig.~\ref{fig:msi_hetero}iv).

Aligning with the risk manifold, epithelial signatures capturing fetal-like and metaplastic states showed positive risk correlations. These included fetal progenitor and epithelial fetal signatures, YAP-driven regenerative and mechanotransduction-related programs. Conversely, signatures reflecting preservation of canonical colonic epithelial identity, including LGR5-positive crypt stem cell programs, ATRX-associated lineage-fidelity signatures and adult intestinal stem cell signatures, showed negative risk correlations.

Several immune-related signatures also tracked with risk. Immune dysfunction and suppression signatures were positively correlated, including transcriptional programs corresponding to $\mathrm{CD8}^{+}$ T cell dysfunction, such as the fully terminally exhausted $\mathrm{CD8}^{+}$ T-cell program characterised by high inhibitory receptor expression. In parallel, a replication stress signature also showed positive risk association, consistent with high-risk states exhibiting stress-associated transcriptional programs. 

\subsection{\textcolor{black}{INSIGHT adds independent prognostic value beyond clinical variables and established molecular pathways}}\label{sec:multivariable}

\textcolor{black}{To quantify INSIGHT's independent prognostic contribution, we performed multivariable Cox proportional hazards analysis in the PETACC-3 cohort (Fig.~\ref{fig:msi_hetero}v). INSIGHT risk was the strongest outcome predictor (HR 1.51, 95\% CI 1.24–1.85), exceeding all clinical and molecular features.}

\textcolor{black}{Among molecular variables, only a limited subset of pathways retained independent association with outcome. The i3 undifferentiated programme\cite{hong_epithelial_2025} showed a positive association with risk (HR 1.49, 95\% CI 1.13–1.96), while Moscat's Transformed Goblet Cells, TGC, (HR 0.73, 95\% CI 0.56–0.94) and Chronic CD8 Activation (HR 0.66, 95\% CI 0.50–0.87) showed inverse associations. In contrast, the majority of evaluated pathway-level features ($n=215$), including established programmes such as i3 metaplasia, did not retain significance in the multivariable model.}

\textcolor{black}{This indicates that INSIGHT captures a substantial proportion of prognostic information encoded in known molecular pathways directly from histology, while leaving a limited set of complementary signals that provide additional stratification value. Incorporating these signals yields a consistent improvement in prognostic performance, increasing the C-index from $0.68$ to $0.74 \pm 0.03$.}

\subsection{\textcolor{black}{Single-cell reference mapping localises INSIGHT-associated risk programmes to fetal-like epithelial and immunosuppressive myeloid states}}
To identify the cellular contexts of genes whose expression correlates with INSIGHT-predicted risk, we mapped their expression patterns onto annotated single-cell transcriptomic reference datasets~\cite{chu2024integrative}. Within epithelial cell populations, genes such as L1CAM, CEACAM5/6 were highly expressed within fetal-like populations. Interestingly, in the stromal compartment we observed non-random associations of high-risk genes in $\mathrm{SPP1}^{+}$ tumour-associated macrophages (TAMs) and $\mathrm{LAMP3}^{+}$ dendritic cells. CD3e and CD8a were enriched in low-risk groups, consistent with the association between tumour-infiltrating T cells and improved outcome.  

Together, these findings associate high predicted risk with epithelial fetal-like reprogramming and myeloid-driven immune suppression, and lower-risk states with adaptive immune infiltration, directly aligning the single-cell projections with the epithelial, myeloid, and adaptive immune axes of the risk manifold.

\begin{figure}
    \centering
    \includegraphics[width=0.7\linewidth]{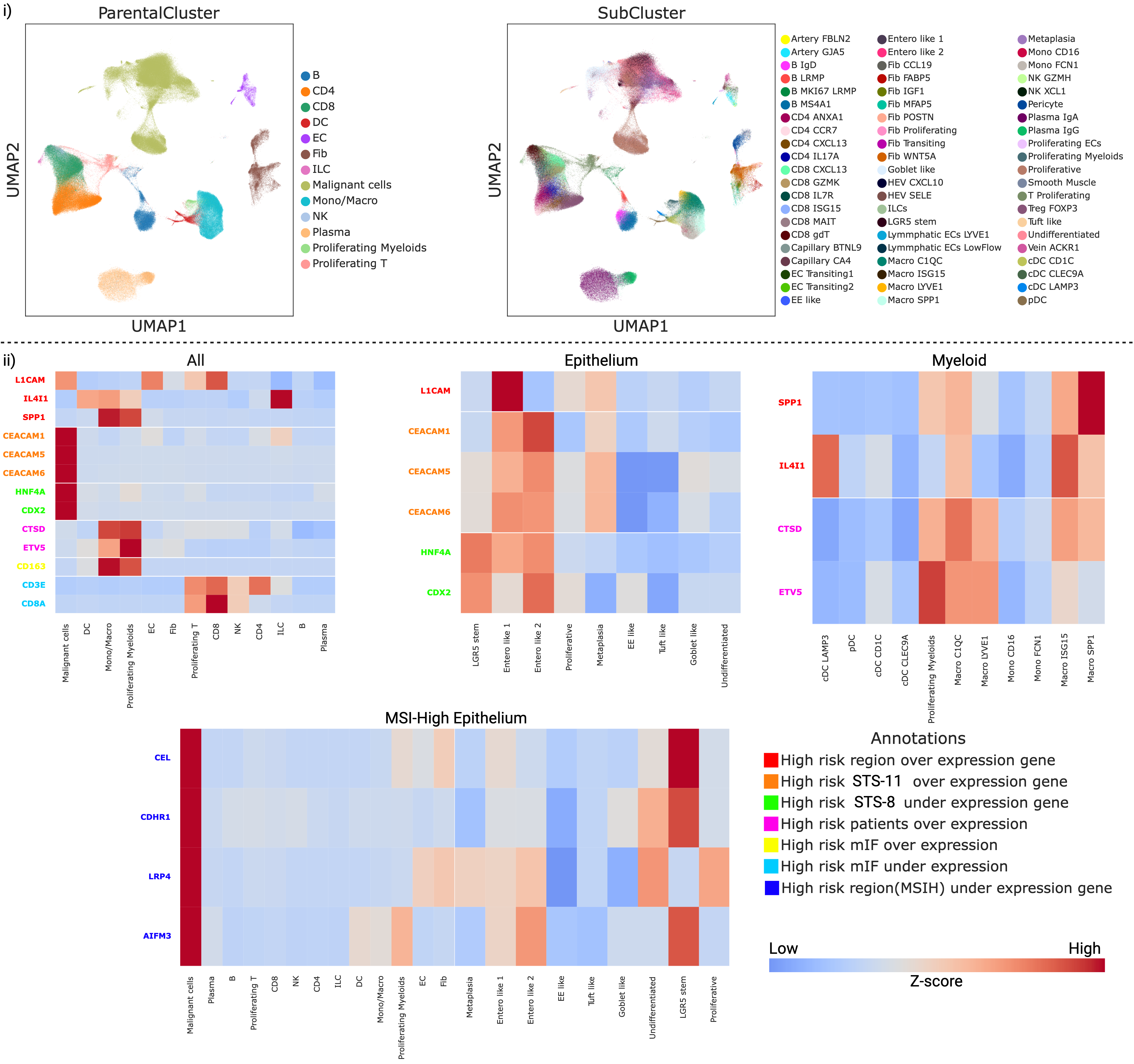}
    \caption{Cell-type interpretation of genes and proteins identified in this study based on single-cell RNA sequencing. In i) on the left, the UMAP shows the integrated single-cell dataset colored by broad cell types; on the right, the same dataset is visualized with fine-grained cell type annotations. In ii) the heatmaps display the average expression levels of genes and proteins identified by INSIGHT as high-risk across different cell populations: from overall broad cell types (upper left) to epithelial subtypes (upper middle) and myeloid cells (upper right), and finally to epithelial cells from MSI-High samples (bottom). We observed non-random expression patterns of these genes in both epithelial cells and myeloid cells within the stromal compartment. Gene colours correspond to their discovery category, while z-scores (calculated from average expression per group and scaled by cell type) indicate relatively high or low expression in each cell type.}
    \label{fig:cell_context}
\end{figure}

\section{Discussion}
Despite advances in image-based risk prediction for stage II/III CRC, the biological basis of these models remains poorly understood. Although deep learning approaches can extract prognostic signals from WSIs, many studies prioritise prediction over interpretability~\cite{skrede2020deep,kleppe2022clinical}. Thus, the mechanisms underlying image-derived risk remain unclear.

INSIGHT addresses two central questions: whether clinically actionable prognostic information can be derived directly from routine H\&E histology, and what biological programmes underlie this risk. Stage II/III disease represents a decision-critical setting in which pTNM staging, while used clinically, provides limited resolution of outcome heterogeneity. Our results show that prognostic signals can be robustly extracted from WSIs and localised within tissue via spatially resolved risk maps. Across internal and external validation on independent cohorts, including the PETACC-3 trial, INSIGHT consistently outperformed pTNM staging and contemporary multiple-instance learning models. 

INSIGHT's performance was strongest in MSI-High, female, and younger patients, groups known to exhibit distinct morphological characteristics. For example, females are often associated with BRAF mutations and serrated pathways with strong histomorphological correlates~\cite{guitton2023artificial}, while MSI-High tumours frequently harbour BRAF mutations, arise through serrated pathways, and exhibit increased immune infiltration~\cite{sadien2024genomics}. Similarly, early-onset CRC is often characterised by aggressive pathological features such as poor differentiation~\cite{stigliano2014early}. These characteristics likely enhance the detectability of histological prognostic signals. Conversely, performance was more modest in stage II disease, consistent with its relative biological homogeneity and lower event rate. 

\textcolor{black}{Importantly, in PETACC-3 multivariable analysis, INSIGHT emerged as the strongest predictor of outcome, exceeding standard clinicopathological variables and most molecular pathway features. Notably, many established programmes did not retain independent significance, indicating that their prognostic signal is largely encoded within histology-derived risk. This includes metaplastic and regenerative epithelial states, suggesting that routine tissue architecture captures a substantial fraction of canonical CRC biology. Even well-established high-risk programmes such as i3 metaplasia did not remain independently significant in the multivariable setting.}

\textcolor{black}{Concurrently, a limited pathway subset remained significant, highlighting complementary biology not fully represented in histology. In particular, the i3 undifferentiated epithelial state (HR 1.49 [1.13–1.96]) reflects a highly plastic phenotype along the stem-to-metaplastic trajectory, characterised by colonic lineage identity loss and activation of stress programmes. Such heterogeneous cellular states may not consistently manifest as stable tissue architecture. Similarly, the chronically sensitised CD8 T-cell signature (HR 0.66 [0.50–0.80]) captures a mixed activation–exhaustion state that may correspond to ambiguous morphological patterns.}

\textcolor{black}{These findings suggest that prognostic information can be decomposed into a dominant spatial component, recoverable from histology, and a residual functional component captured by molecular profiling, reflecting biological processes that do not consistently manifest as stable morphological patterns. Importantly, incorporating these signals with INSIGHT further improves prognostication (C-index = $0.73 \pm 0.04$), demonstrating a clear path toward multimodal models. INSIGHT therefore functions as a clinically applicable, histology-based integrator of prognostic biology that surpasses current standards of care, while providing a principled framework for combining spatial and molecular information to enhance prognostication.}

INSIGHT's key advantage over pTNM staging is the ability to localise prognostic signal, providing a direct link to underlying tumour biology. Spatial risk maps revealed that high-risk regions consistently align with established adverse histomorphological features, including dedifferentiation and increased stromal content~\cite{knight2024glasgow,backes2018histologic}, supporting INSIGHT's biological validity. 

Beyond recapitulating known features, quantitative morphometric analysis identified increased nuclear solidity and circularity as robust risk correlates. These provide an interpretable and automatable proxy for epithelial state, reflecting chromatin compaction and architectural disruption associated with fetal-like transcriptional programmes. \textcolor{black}{This is consistent with recent mechanistic studies demonstrating that nuclear deformation can induce fetal-like cellular states through YAP-mediated transcriptional reprogramming~\cite{bertillot2025colorectal}. This highlights that INSIGHT-derived risk is not only statistically robust but also grounded in mechanistically meaningful tissue phenotypes, linking image-derived risk to fundamental epithelial plasticity processes.}

INSIGHT's spatially resolved nature enabled direct integration with multimodal data. By aligning patch-level risk maps with spatial transcriptomics, mIHC, and mIF measurements, we treated localised risk as an outcome surrogate, allowing molecular features to be interrogated in their spatial context. This revealed that prognostic biology is organised into coordinated spatial tissue states, rather than isolated molecular events, which we formalise into a risk manifold.

Axis 1 corresponded to an undifferentiated, fetal-like epithelial state, consistent with invasive, highly plastic tumour populations marked by loss of colonic lineage and treatment resistance~\cite{moorman2025progressive}. This compartment is a major therapeutic focus, with strategies targeting CEACAM5/6 and LGR5~\cite{kopetz2025precemtabart,kogai2025broad}. However, epithelial state alone does not fully explain prognostic variation. Instead, two additional axes capture complementary stromal and immune risk contributions, indicating that adverse outcomes arise from coordinated interactions between tumour-intrinsic and microenvironmental processes.

Axis 2 reflected a myeloid-driven immunosuppressive microenvironment. High-risk regions were characterised by enrichment of suppressive myeloid populations, including $\mathrm{SPP1}^{+}$ tumour-associated macrophages and $\mathrm{LAMP3}^{+}$ dendritic cells, consistent with a stromal context that promotes tumour progression and dampens effective immune responses. This highlights the contribution of innate immune regulation to prognosis, complementing tumour-intrinsic epithelial programmes. While targeting the CSF1–CSF1R axis to modulate tumour-associated macrophages is an active therapeutic strategy, the prominence of $\mathrm{SPP1}^{+}$ TAMs identified here may explain limited clinical efficacy~\cite{ltz_therapeutics_inc_phase_2025}.

Axis 3 captures adaptive immune dysfunction. High-risk regions were associated with reduced cytotoxic T-cell activity alongside signatures of T-cell exhaustion, including diminished $\mathrm{CD8}^{+}$ infiltration and increased inhibitory programmes. This indicates that impaired adaptive immune surveillance, rather than the simple absence of immune infiltration, contributes to adverse outcomes. Although immune checkpoint blockade targets this axis, the coexistence of epithelial plasticity and myeloid-mediated suppression suggests that adaptive immune dysfunction operates within a broader, multi-compartmental framework of tumour progression~\cite{tian_global_2021}.

Although pharmacologic interventions exist for each axis, current therapeutic strategies do not address their coordinated interaction. Our results suggest that therapeutic approaches that target these axes concurrently may show improved efficacy.

\textcolor{black}{Notably, many risk manifold features show concordance with independently derived mechanistic models of epithelial state regulation~\cite{buissant2026emergence}. This includes organoid systems, which have demonstrated that epithelial plasticity is governed by a conserved regulatory axis defined by YAP/AP1-driven regenerative programmes~\cite{perez2025self} and opposing differentiation signals mediated by RXR/PPAR activity,\cite{mzoughi_oncofetal_2025} modulated by stromal and mechanical cues~\cite{kirino2026fetal}. This aligns closely with our observations: YAP/AP1-associated regenerative and fetal-like programmes (L1CAM) are enriched in high-risk regions, whereas differentiation-associated signals, reflected in STS-8 and related epithelial identity programmes, are diminished (Axis 1). Concurrently, stromal enrichment and myeloid-associated features, including SPP1$^{+}$ macrophages, are consistent with the microenvironmental signals known to induce and stabilise these epithelial states (Axis 2). Importantly, while organoid and experimental systems robustly capture epithelial and stromal-driven plasticity, they typically lack a fully developed adaptive immune compartment. The emergence of a distinct immune dysfunction axis (Axis 3) in our patient-derived data, therefore, extends these mechanistic models, capturing additional layers of biology not accessible in reductionist systems. Recent spatial and mechanistic studies also support this interpretation, showing that CRC progression is shaped by spatially organised malignant-immune hubs, SPP1$^{+}$ macrophage niches, CDX2/HNF4A-dependent lineage maintenance, and stromal mechanical cues that induce fetal-like epithelial reprogramming~\cite{matusiak2024spatially}. Together, this convergence between unbiased histology-derived patterns and controlled mechanistic models provides strong orthogonal validation of the risk manifold and supports its interpretation as a biologically grounded disease progression representation.}

To further interpret the risk manifold, we examined molecular features most consistently associated with risk. Spatial transcriptomic analysis revealed recurrent gene programmes linked to prognosis across diverse patient contexts, including high-risk fetal-like genes (L1CAM, IL4I1, SPP1) and low-risk differentiated epithelial markers and immune states (CXCL14, CES2, PLCE1)~\cite{joanito2022single}. L1CAM, in particular, was enriched in iCMS2 Entero\_like\_1 and iCMS3 Metaplasia epithelial populations, both previously associated with poor survival~\cite{hong_epithelial_2025-1}, reinforcing the link between epithelial plasticity and adverse outcome.

Building on this gene-level analysis, we examined the transcriptional programmes that underlie the epithelial risk axis. We analysed Spatial Transcriptomic Signatures (STS), which represent data-driven, co-expressed gene modules capturing coordinated spatial expression risk patterns. Multivariate STS analysis identified two primary signatures explaining localized prognostic variation, with STS-8 contributing the largest effect. 

STS-8, defined by underexpression of canonical intestinal lineage markers, was consistent with patterns observed in iCMS3 Metaplasia and iCMS2  Entero\_like\_1 epithelial cells~\cite{noauthor_epithelial_2025}. The deregulation of CDX2, a central feature of STS-8, is widely associated with invasive phenotypes. Direct CDX2 targets, including HNF4A, which maintains colonic identity, and CHDR5, which modulates tumour formation via adhesion were co-downregulated~\cite{cammareri2025loss}. Dedifferentiation suggested by STS-8 is accompanied by L1CAM expression, indicating a shift toward fetal-like states and a loss of normal epithelial constraints~\cite{bodmer2025cdx2}. STS-8 was highly activated in all BRAF-mutated and two BRAF-like cases. Similarly, KRT20, an intestinal differentiation marker was diminished, consistent with known links to BRAF-mutant tumours and aggressive phenotypes~\cite{spisak2024identifying}. This highlights a mechanistic link between STS-8 activation and BRAF-driven CRC. STS-8 also reflects loss of additional intestinal functions: CDX2-dependent transmembrane mucins MUC12 and MUC17, aligning with previous observations of poor prognosis and epigenetic silencing~\cite{matsuyama2010muc12}. RORC and CXCL14, genes associated with invasion and hypermethylation, were similarly reduced, further supporting aggressive biology connections~\cite{wang2023data}. A novel link with CES2 can be observed, typically linked to chemoresistance, and COL17A1, which modulates drug sensitivity via an mTOR-Akt scaffold~\cite{lin2024collagen}. This may indicate a unique CRC phenotype where canonical lineage loss occurs alongside functional rewiring of drug response pathways. Additional genes, including SULT1B1, RBP4, NXPE4, and MYH14, were also strongly downregulated, highlighting potential genes for future study~\cite{bryan20204}. For example, activation of MYH14 by 4-HAP inhibited metastasis in mice, and its importance to STS-8 highlights the urgency for human study. Overall, STS-8 supports established biological functions while revealing previously unrecognised co-regulated genes and pathways that may contribute to aggressive CRC phenotypes and therapeutic responses.

STS-11 emerged as the second-highest risk contributor, characterized by CEACAM family members overexpression (CEACAM1/5/6). STS-11 appears in both fetal-like/metaplastic populations and lineage-retaining enterocyte-like cells, indicating that CEACAM expression alone does not uniquely define high-risk epithelial states and should be interpreted in an epithelial differentiation context. Most patients with high STS-11 activation harboured TP53 and APC mutations, whereas low activation was observed in BRAF- or KRAS-mutant cases. CEACAM genes are well-established CRC progression drivers: CEACAM6 promotes chemoresistance, while CEACAM1, shows pro-metastatic properties~\cite{chu2023aldolase}. High CEACAM5 tissue expression has been correlated with worse survival~\cite{aldilaijan2023clinical}. Other co-expressed genes in STS-11, including S100P, UBE2C, and AZGP1, further support a poor prognosis phenotype through enhanced migration and metabolic adaptation~\cite{jalali2024ube2c}. STS-11 also points to curated Wnt signaling: ASCL2, a direct Wnt target controlling stem cell fate, is upregulated alongside downregulation of RSPO3, suggesting an optimized Wnt/RSPO balance potentially reinforced by mutational status~\cite{li2024rnf43}. MEIS2, implicated in CRC invasion, is downregulated, though its integration with STS-8 remains to be explored~\cite{wan2019meis2}. CFTR, a tumour suppressor, is co-upregulated with CEACAMs, potentially marking a secretory subtype overlaid with oncogenic programs~\cite{spelier2024cftr}. Overall, STS-11 highlights a distinct CRC risk route, combining proliferative, Wnt-responsive, and metabolic programs. While validating known CEACAM roles, it also uncovers coordinated metabolic and transcriptional patterns that may underpin aggressive disease, warranting further investigation.

Beyond epithelial programmes, the risk manifold also captures distinct stromal and immune contributions. In the myeloid compartment, high-risk regions were characterised by $\mathrm{SPP1}^{+}$ tumour-associated macrophage enrichment, a population associated with immunosuppression phenotypes~\cite{matusiak2024spatially}. Similarly, IL4I1 in $\mathrm{LAMP3}^{+}$ dendritic cells reflects a metabolic immune checkpoint that suppresses antitumour T-cells, further contributing to an immunoregulatory microenvironment~\cite{you2024lymphatic}. Analysis of mIF and mIHC data confirmed that T-cell (CD3e, CD8a) and epithelial (CDX2) markers were enriched in low-risk regions, whereas macrophage markers (CD68, CD163) predominated high-risk tissue, reinforcing the link between myeloid-rich microenvironments and adverse outcomes. Together, these findings show the risk manifold integrates epithelial, myeloid, and immune programmes into a coherent spatial framework of prognostic biology.

More fine-grained patient-level analyses suggest that the relationship between molecular programmes and risk is context-dependent. Notably, patients with a Tubulovillous Adenoma (TVA) precursor show distinct gene correlation patterns, potentially reflecting differences in tumour origin and evolutionary trajectory. LGR5 showed patient-specific risk associations, consistent with its dual role in CRC prognosis~\cite{noauthor_rspo2lgr5_nodate}. This likely reflects context-dependent regulation of the LGR5/WNT axis and tumour microenvironmental interactions, including epigenetic control of transitions between LGR5$^{+}$ and LGR5$^{-}$ states~\cite{perez2025self}. Similarly, elevated EGFR expression in IMU004, a high-grade dysplasia precursor, highlights an alternative oncogenic route associated with therapeutic resistance, consistent with EGFR-targeted therapy escape mechanisms~\cite{poh2022therapeutic}. Together, these findings indicate that prognostic programmes are shaped by tumour-specific contexts, reinforcing the need to interpret risk within a framework that accommodates biological heterogeneity rather than relying on uniform biomarkers. However, given the limited size of the IMMUcan cohort, these observations require further validation.

Beyond localized patterns, patient-level risk analyses revealed heterogeneity within MSI-High tumours. Low expression of CEL, CDHR1, LRP4, and AIFM3 was associated with higher predicted risk, whereas single-cell mapping showed these genes are enriched in LGR5\_stem epithelial cells, typically linked to lower-risk states compared with Entero\_like\_1 and Metaplasia \cite{hong_epithelial_2025}. This suggests MSI-High tumours retaining an LGR5-stem-like epithelial program may represent a biologically distinct, lower-risk subgroup. Further immune marker analysis indicated that high-risk MSI-High cases exhibit an overactive, CD8-exhausted immune microenvironment, consistent with chronic inflammation observed alongside fetal-like states~\cite{goto2024sox17}. These highlight MSI-High heterogeneity and warrant further validation.

\textcolor{black}{In summary, INSIGHT outperforms traditional patient stratification methods and provides independent prognostic value beyond standard clinicopathological variables. Unlike pTNM staging, it generates localised risk predictions, enabling spatially resolved interrogation of tumour biology. Importantly, INSIGHT recovers established prognostic features while uncovering novel associations, supporting both its validity and biological relevance. Cross-modal analysis reveals that these signals are organised along a coherent epithelial–immune risk manifold that explains INSIGHT's predictions and provides actionable biological insight. High-risk regions are characterised by fetal-like or regenerative epithelial programmes and enrichment of tumour-associated myeloid cells, whereas low-risk regions are associated with adaptive immune function.}

This indicates that prognosis emerges from coordinated interactions between epithelial and microenvironmental compartments, rather than epithelial state alone. We hypothesise that chronic inflammation sustained by myeloid populations may promote or stabilise fetal-like epithelial reprogramming. This highlights the need for therapeutic strategies that target both epithelial plasticity and the supporting stromal and immune context.

Key limitations include the relatively small spatial transcriptomics datasets, and the discovered manifold may be biased by available markers and not capture the full prognostic landscape. Future work will validate findings in larger cohorts and assess ctDNA integration~\cite{nakamura2024554p}.

\section{Ethic Statements}
The IMMUcan cohort ethics reference is S63391. The IMMUcan project has received funding from the Innovative Medicines Initiative 2 Joint Undertaking under grant agreement No 821558. This Joint Undertaking receives support from the European Union's Horizon 2020 research and innovation programme and EFPIA. IMI.europa.eu

The Janssen cohort ethics reference is S62294 and S64121.

\section{Conflicts of Interest}
S.T. is supported by a BOF-Fundamental Clinical Research mandate (FKO) from KU Leuven (DOC/CM21-FKO-02) and KULeuven internal research project C24M/22/055.

GR has no COI in relation to this project. 

\bibliographystyle{vancouver}
\bibliography{lancet-sample}

\newpage{}

\section{Methods}

\subsection{WSI datasets with survival information}
We collected colorectal cancer (CRC) samples from three cohorts containing routine Haematoxylin and Eosin (H\&E) stained Formalin Fixed Paraffin Embedded (FFPE) WSIs. These included data from: 1) The Cancer Genome Atlas (TCGA), 2) SR386 cohort with survival information from SurGen~\cite{myles2025surgen} and the 3) randomized phase III trial PETACC-3~\cite{van2009randomized}. After filtering for patients who were labelled as stage II or III we ended up with 1833 patients, with 342 from TCGA, 336 from SR386, and 1155 from PETACC-3. To rigorously evaluate INSIGHT we performed 4-fold Cross Validation with publicly available TCGA and SR386 data. The folds were patient-aware to ensure we tested on completely unseen cases and stratified on event (patient death) to ensure an even distribution of labels in the train, test split. PETACC-3 was then used as an external test set. In this work, we use Overall Survival (OS) censored at 3 years as the endpoint for training and testing.

\subsection{Patch-level datasets for tissue phenotyping}
To develop a model for segmentation of different types of tissue, we collected data from 4 sources that contained patch level annotations of tissue regions from Stage I-IV colorectal cancer (CRC) samples. These were labelled into one of the following 8 classes: Tumour (TUM), Stroma (STR), Lymphocyte (LYM), Mucosa (MUC), Normal Gland (NORM), Muscle (MUS), Necrosis/Debris (DEB) and Adipose (ADI). All patches were scanned at 0.5 micrometers-per-pixel (MPP) and were scaled to 224 x 224 pixels. The first dataset was NCT-CRC-HE which provided 95,766 patches (after removing the background class)~\cite{armand2025automated}. Secondly, we utilised TSR-CRC which provided 180,880 patches~\cite{armand2025automated}. Thirdly, we utilised the CFHCD dataset that provided 1866 tumour and stromal patches~\cite{carvalho2025ai}. Finally, we utilised the patch-level dataset from Cerberus which provided 118,115 patches from all 8 classes~\cite{graham2023one}. Thus, in total we ended up with 396,632 patches.
To train the patch-level classifier we used a Vision Transformer (ViT) model pre-trained on ImageNet-21k. This model was trained for 30 epochs with a batch size of 32, learning rate of  $5\cdot 10^{-5}$ and early stopping after 10 epochs. To reduce GPU memory cost, we used fp16 precision. The resulting model achieved a balanced accuracy on the test set of 91.28 across all 8 tissue types.

\subsection{WSI Pre-processing and patch representation}
For each WSI we first identify its viable tissue regions via GrandQC which performs tissue detection as well as artefact removal~\cite{weng2024grandqc}. Artefacts were defined as either: pen markings, tissue folds or dark spots. We chose to use this model as both TCGA and PETACC-3 have a large quantity of artefacts which may act as confounding factors and thus must be removed. All regions from the detected tissue are then passed into the tissue classifier that assigns each 224 by 224 patch at 0.5 MPP one of 8 tissue labels. It was important to use the initial time efficient GrandQC tissue mask to reduce the tissue area our segmenter needs to analyse as well as to filter out regions the model does not know how to interpret (background and artefacts). After this, only patches that were predicted as either TUM, STR or LYM were kept, producing a new binary mask for each WSI with a score of one for kept tissue area and zero otherwise. Since WSIs at full resolution can be very large ($100,000 \times 100,000$ pixels) and cannot fit into GPU memory, we apply the previously generated masks and then tile each WSI into patches of size $512 \times 512$ at a spatial resolution of 0.5 MPP. Patches capturing less than 40\% of tissue area (mean pixel intensity above 200) are discarded, and the rest of the patches are used. For each patch in a WSI we obtain its d=1024-dimensional feature representation by passing it through the UNI feature encoder~\cite{chen2024towards}. We do not apply colour normalization before patch feature extraction, as previous work has found it to have limited benefit~\cite{keller2023tissue}. Further, data augmentations have been applied during training of the encoder, thus it should be robust to stain variation.

\subsection{Graph Construction}
Mathematically, a  graph for a WSI is defined as $G_k \equiv (V_k,E_k) $ where $V_k$ is a set of vertices and $E_k$ is a set of edges, where $\{i,j\} \in E_k$ denotes an edge between nodes $i$ and $j \in V_k$. $V_k$ describes the set of all valid patches in WSI $k$. For each vertex we have an associated 1024-dimensional feature vector as extracted by the UNI feature encoder~\cite{chen2024towards}. An edge is defined based on Delaunay triangulation which connects neighbouring patches based on their spatial closeness while avoiding overlapping connections. This approach helps ensure that each patch is linked to the most relevant nearby patches, capturing the tissue's structure in a meaningful and efficient way.  To avoid connecting patches that are too far apart, we apply a distance threshold of 4000 pixels (2 millimetres).

In CPath we typically have multiple WSI per patient but only a single patient-level outcome label. Thus, like a pathologist, we need a way to aggregate information from all WSIs to produce a single patient-level prediction.  This can be done manually via different aggregation methods such as taking the mean, max, or median across individual WSI predictions.  However, there is no obvious single best choice for all patients. Thus, we instead chose to construct a patient-level graph composed of all the patient's WSI-level graphs. Mathematically, for a patient $p$ with $N_p$ WSIs, $W \coloneqq  \{G_1,\ldots,G_{N_p}\}$, we can define its graph as $G^p \equiv (V^p,E^p)$. Here $V^p = \coprod_{m \in W}V_m$ and $E^p = \coprod_{m \in W}E_m$ are defied as disjoint union of vertices and edges respectively across all WSIs of patient $p$. By following this definition, we make the aggregation of WSI-level information automatic since the model can see all the information at once and decide for itself what is the most important to prognosis.

\subsection{Patient-level Graph Neural Network}
After creating a patient-level patch graph, $G^p \equiv (V^p,E^p)$, we can use this as input to a Graph Neural Network (GNN). Each Transformer layer $l=(1,2,3)$ updates each node, $i$, representation by aggregating information from its neighbours, $\mathcal{N}(i)$ and itself via the rule: 
\begin{equation}
h^{l}_i = \beta_i \mathbf{W}_1 h_i^{l-1} +
(1 - \beta_i) \left(\sum_{j \in \mathcal{N}(i)}
\alpha_{i,j} \mathbf{W}_2 \vec{h}_j^{ l-1} \right)
\end{equation}
where $\mathbf{W}_1,\mathbf{W}_2$ depict learnable weights and $\alpha_{i,j}$ depicts the learnable attention coefficient between the current node $i$ and its neighbours $\mathcal{N}(i)$, allowing us to weigh the contribution of neighbours. Additionally, $\beta_{i}\in(0,1)$ is a learnable parameter that decides what proportion of information should come from the node itself and how much from its neighbourhood.  After applying all layers, we end up the final node-level feature embedding $r_i$ which can be directly used to show patch-level risk. To obtain the final patient-level risk score, $f(G^p)\in \mathbb{R}$ we perform global attention pooling:
\begin{equation}
f(G^p) = \sum_{i \in V^p} \frac{exp(\omega(r_i))}{\sum_{j \in V^p} exp(\omega(r_j)) } \cdot \phi(r_i)
\end{equation}
where $\omega$ and $\phi$ are Multi-Layer Perceptrons. This allows INSIGHT to learn to weigh patches based on their varying importance to overall patient-level risk score, essentially ensuring the model focuses on the most important regions.

\subsection{Understanding INSIGHT outputs}
INSIGHT produces three complementary outputs, each with prognostic value. First, INSIGHT generates a localized patch-level risk score, normalized to lie between 0 and 1, where 0 indicates lowest risk and 1 indicates highest risk. Normalization is performed using min-max scaling based on the 2nd and 98th percentiles of patch-level risk scores from the training set. This strategy reduces sensitivity to outliers and ensures that patch-level risk scores are directly comparable not only across patients, but also across datasets, including external cohorts.

Second, INSIGHT produces a patient-level risk score that summarizes overall risk for each patient by aggregating information across all patches from all whole-slide images associated with that patient. Importantly, this aggregation is learned by the model, allowing INSIGHT to weight patches according to their relative importance in determining patient-level risk, rather than treating all patches equally. The resulting score provides a relative ranking of patients, with higher values indicating higher predicted risk.

Third, INSIGHT outputs a binary patient-level risk stratification, assigning each patient to either a low- or high-risk group. This stratification is used for Kaplan-Meier survival analysis. Patients with risk scores above a predefined threshold are assigned to the high-risk group, while those below the threshold are assigned to the low-risk group. The threshold is optimally determined using the training data only, via a grid search to identify the value that maximizes survival separation as measured by the log-rank test p-value. No test data are used in defining this threshold.

\subsection{Multimodal datasets for prognostic factor discovery}
We collected 3 datasets of multi-modal data that contained a unique data modality as well as matching H\&E WSIs. This means we can use INSIGHT to generate patch-level H\&E based predictions which can then be cross-referenced with variables from orthogonal data modalities in the same regions.  Firstly, we use the Janssen cohort which contains 236 H\&E WSIs ($n=80$) with matching multi-Immunohistochemistry (mIHC) WSIs. These mIHC WSIs were stained with CDX2, CD8, MUC2, MUC5 and Haematoxylin.  Secondly, we used the publicly available Orion dataset which contains 27 H\&E WSIs ($n=27$) with matching multi-Immunofluorescence (mIF) WSIs measuring 19 protein markers.  Finally, we used the IMMUcan dataset of 15 patients which have spatial transcriptomics measuring expression of 422 genes using 10x Genomics. These genes were selected using the standard Xenium colon panel as well as 100 genes from previous single cell work on CMS/iCMS.

\subsection{Multimodal WSI Registration}
To utilise multimodal datasets we first need to register the H\&E WSIs with the additional WSI data modalities. Janssen and IMMUcan were registered via DeeperHistoReg, a tool that performs automatic robust non-rigid registration~\cite{wodzinski2024deeperhistreg}. DeeperHistoReg has been modified to work a wide range of WSI formats and data modalities that could not previously be processed, by incorporating a new WSI reader into the pipeline. For the Janssen cohort we directly registered mIHC to H\&E whereas for spatial transcriptomic data we registered the H\&E to a DAPI WSI which was already registered to spatial transcriptomic data. Orion WSIs were previously registered using PALOM~\cite{lin2023high}.

\subsection{Generating patch-level expression from cell-level}
Our goal was to correlate patch-level risk with spatial transcripto mics and mIF expression. However, since mIF and spatial transcriptomics measure expression at a cell level, we need a way to aggregate the expression of all cells in each patch into a single expression value. The simplest way is to take a patch p which contains a set of cells $C$ and take the mean across cells producing a patch expression score  $p_s = \frac{1}{|C|}\sum_{i \in C} c_i$ where $c_i$ is the expression of a certain gene or protein. However, this approach may not always work as raw gene expression values across cells within a patch vary widely in scale and distribution, making direct comparisons across genes unreliable. Additionally, because patches contain differing numbers of cells, it is difficult to aggregate expression values in a scale-invariant and biologically meaningful way. To address this, we learn a gene-specific sigmoid transformation that maps each cell-level expression to a standardized $[0, 1]$ scale. This transformation preserves the ordering of expression values while learning a soft threshold and sharpness per gene, capturing "switch-on/switch-off" behaviour. Averaging these transformed values across all cells in a patch yields a normalized, risk-aligned gene activation score that enables interpretable pooling of cellular expression and consistent comparison across patches. This can be expressed as $p_s = \frac{1}{|C|}\sum_{i \in C} \sigma (a \cdot (c_i - b))$  where $\sigma$ is the sigmoid function and $a,b \in R$ are parameters that scale the data. By modifying the values of $a$ we define the quantity of a gene/protein needed to activate it, a higher a means you need less of it. The value of $b$ defines at what quantity does a protein/gene get activated. To find the best value of $a$ and $b$ we perform grid search, selecting the combination that produces the highest spearman correlation with patch-level risk across all patients and patches. By automating the selection of $a,b$, we essentially cycle through different aggregation function where as $a \rightarrow0$ we perform simple mean aggregation and as a get larger we tend towards taking lower quantiles, as even a tiny amount of a gene/protein causes the sigmoid to push the value to $1$ (on).

\subsection{Spatial Transcriptomic Gene Modules}
After defining patch-level spatial transcriptomic gene expression, we can perform signature discovery to uncover coherent sets of genes whose local expression patterns contribute to tissue-level risk in a transparent and biologically grounded manner.  This is done because genes often act in coordinated groups rather than in isolation. To this end, we utilise CoRex to group genes into 20 binary Spatial Transcriptomic Signatures with each signature comprising a set of genes whose expression is co-dependent. This means for each patch, we can infer whether a signature is active (1) or inactive (0), with the constraint that signature activation is positively associated with increased risk for interpretability. Importantly, although a signature's activation reflects a risk-enhancing pattern, individual genes within a signature may be positively or negatively associated with its activation, capturing natural co-expression diversity. To run CoRex we first construct a matrix $Z \in R^{N \times J}$, where each row corresponds to the sigmoid-transformed patch-level gene activations $z_i^p \in R^J$ for a patch with $J$ genes. We then apply the Lambert's W function to each gene in $Z$. This is because many genes follow a heavy-tailed distribution after scaling via a sigmoid function however CorEx assumes normal distributed genes thus we need to perform this transformation. With this we can run CorEx which will allow us to express each patch via a 20 variable binary vector of the form: $g_i^p=[g_{i,1}^p,\cdots,g_{i,20}^p]\in\{0,1\}^{20}$.

Finally, we need to quantity the association of each signature with predicted risk. This can be done by fitting a linear regression model (OLS) using binary signature activations to predict patch-level risk. The resulting coefficients, quantify the independent contribution of each signature to the risk score, offering a clear and interpretable link between spatial gene programs and H\&E based risk predictions. An OLS coefficient  $\beta_t$ measures the expected increase in predicted risk when signature $t$ is active ($g_i^p,t=1$), assuming all other signatures remain fixed. A positive  $\beta_t$  directly quantifies the risk-increasing effect of signature $t$. 

\subsection{mIHC stain deconvolution}
The Janssen mIHC whole slide images (WSIs) were simultaneously stained with haematoxylin, CDX2 (pink), CD8 (brown), MUC2 (yellow), and MUC5 (green). To quantify the concentration of each stain and relate it to localized risk predictions, we needed a method to separate these stains. The simplest solution would be to generate a pure stain matrix for colour deconvolution. 

However, this performs poorly due to the high similarity between stain colours, and the fact that we are attempting to map five stain channels into three RGB channels, which is mathematically underdefined. To address this, we trained a custom U-Net which aims to reconstruct the WSI using a learnable stain-matrix based decoder. This approach produces more robust stain separation by leveraging not only colour information but also morphological cues from local image regions. We used a combination of different losses to ensure the model: can reconstruct the image (L1 and perceptual loss), prefers pure stains (L1 loss in concentration space with per-pixel extropy based loss on concentrated maps) and shows consistently with the pure stain matrix (L1 loss between learned stain matrix and pure one).

Using this trained stain matrix we can generate a patch-level concentration map that tells us the per-pixel concentration of all five stains. Similarity to generating patch-level expression from cell-level, we can use a learnable sigmoid function to identify the best way to combine these per-pixel stain concentrations to a single patch-level stain concentration which can then be correlated with localised risk predictions.

\subsection{Cell-Type analysis}
To validate our identified high-risk genes and signatures in our analysis, we utilized previously described tumor epithelial labels and an annotated single-cell transcriptomic dataset from approximately 200 donors~\cite{hong_epithelial_2025-1,chu2024integrative}.  To achieve label transfer we performed cross-dataset annotation using scPoli, a probabilistic label transfer framework within scArches. Epithelial labels defined in the YH et al. reference dataset were projected onto the epithelial compartment of the Chu et al. dataset using scPoli's default workflow~\cite{hong_epithelial_2025-1,chu2024integrative}. Briefly, scPoli employs a variational autoencoder to construct a shared latent space between reference and query cells, identifies nearest neighbors across datasets, and transfers labels through probabilistic neighborhood voting, thereby capturing annotation uncertainty. The transferred epithelial annotations were then integrated with the native tumor microenvironment (TME) labels of the Chu et al. dataset for joint visualization and downstream analyses. In the heatmap validation, for selected genes, we calculated mean expression per group from the AnnData object. Groups were optionally ordered using a dendrogram, and values were transformed to gene-wise z-scores. The results were visualized as a heatmap with seaborn.

\subsection{Patient-level clinical and molecular data}
Patient-level clinical molecular data used for correlation analyses in this study included bulk RNA-sequencing gene expression data and single-sample Gene Set Enrichment Analysis (ssGSEA) scores for multiple single-cell-derived gene signatures commonly used in colorectal cancer (CRC) research. These provided a list of genes per signatures derived from mouse, human and organoid data. Detailed descriptions of the data sources and gene definitions for each signature are provided in the supplementary Excel file. 

For each gene or pathway, univariate analyses were performed by calculating spearman rank correlation coefficients and corresponding p-values with predicted patient-level risk. To correct for multiple hypothesis testing, the Benjamini-Hochberg (non-negative) procedure was applied. 

\subsection{Multivariable Cox model}
\textcolor{black}{Multivariable Cox proportional hazards modelling was performed in the PETACC-3 cohort. Patients with missing values were excluded prior to analysis, and all continuous variables were standardised to zero mean and unit variance. We initially considered a set of 215 pathway-level molecular features derived from bulk RNA-seq data. To reduce dimensionality and avoid overfitting in the Cox model, feature selection was performed using a LASSO-regularised Cox regression applied to the molecular features only. The top 10 pathways, ranked by the magnitude of their coefficients, were selected for downstream analysis. A final multivariable Cox proportional hazards model was then fit including the selected 10 molecular features together with four predefined clinical variables and the INSIGHT risk score. Hazard ratios (HRs), 95\% confidence intervals, and p-values were obtained from the fitted model. All analyses were implemented in Python using standard survival analysis libraries.}

\subsection{Learning the risk manifold}
We model patch-level disease risk as projections of multimodal morpho-molecular features onto a low-dimensional risk manifold. To estimate this manifold across heterogeneous datasets with systematic multimodal missingness (e.g. spatial proteomic/transcriptomic measurements available only for a subset of patches), we use a biologically structured, data-driven axis model with explicit missingness handling.

For this purpose, molecular or image derived features are grouped into a biologically interpretable axes reflecting dominant biological themes observed in the data (myeloid immunosuppression, epithelial differentiation, and autoimmune suppression). For each axis $a$, we define a feature set $F_a$ with anchored signed weights $w_{fa}^0$ (from biological knowledge and correlation analysis), whose sign encodes the expected direction of influence. The model learns a shrink-only scaling factor $r_{fa} \in [0,1]$ (parameterised as $r_{fa}=1-\sigma(\rho_{fa})$) and forms the effective weight $w_{fa}=w_{fa}^0 \cdot r_{fa}$, preserving directionality while allowing data-driven attenuation of feature contributions. Weights are fixed to zero for features not assigned to axis $a$. Let $x_{pf}$ denote the standardised value of feature $f$ for patch $p$, and let $m_{pf}\in \{0,1\}$ indicate whether that feature is observed. The axis score for a patch $p$ along axis $a$ is computed using a missingness-aware normalised weighted average: $s_p^a=\frac{\sum_{f \in F_a} m_{pf} \cdot w_{fa} \cdot x_{pf}}{\sum_{f \in F_a} m_{pf} \cdot |w_{fa}| + \epsilon}$. Patch-level risk is then predicted by linear regression: $\hat{y}_p=\alpha+\sum_{a=1}^A \beta_a \cdot s_p^a$ with optional interaction terms between axes. All parameters are learned using gradient-based optimisation to minimise the mean squared error between predicted and observed INSIGHT patch risk across all patches, with $L_2$ regularisation on feature-axis weights to stabilise training and prevent overly large contributions from individual features. The resulting feature weights $w_{fa}$ capture the contribution of each feature to axis $a$, while the regression coefficients $\beta_a$  quantify the overall impact of each axis on patch-level risk. Features are assigned to axes without overlap, enforcing structural separation between biological processes. However, axis scores are not constrained to be orthogonal and may remain correlated due to underlying biological coupling.

\subsection{Software, optimisation, and reproducibility}
Our model was developed using: Pytorch, PyTorch Geometric and Python on a NVIDIA DGX A100 with a single NVIDIA Tesla V100 GPU. The GNN was trained for 100 epochs, with a batch size of 32 and optimised with the AdamW algorithm with a learning rate of $5 \cdot e^{-5}$ and weight decay of 0.0005 and early stopping set to 5 epochs. For the learning rate scheduler we used reduce learning rate on plateau with a factor of 0.35, patience of 15 and min learning rate of $2.5\cdot e^{-6}$.To reduce memory usage during training we also utilised fp16 precision. The interactive visualisation was developed using the tile server from TIAToolbox and Bokeh.

\section{Extended Results}

\setcounter{figure}{0} 

\renewcommand{\thefigure}{\arabic{figure}}
\captionsetup[figure]{labelfont=bf, name=Extended Figure}

\setcounter{table}{0}  
\renewcommand{\thetable}{\arabic{table}}  
\captionsetup[table]{labelfont=bf, name=Extended Table}

\begin{figure}
    \centering
    \includegraphics[width=0.7\linewidth]{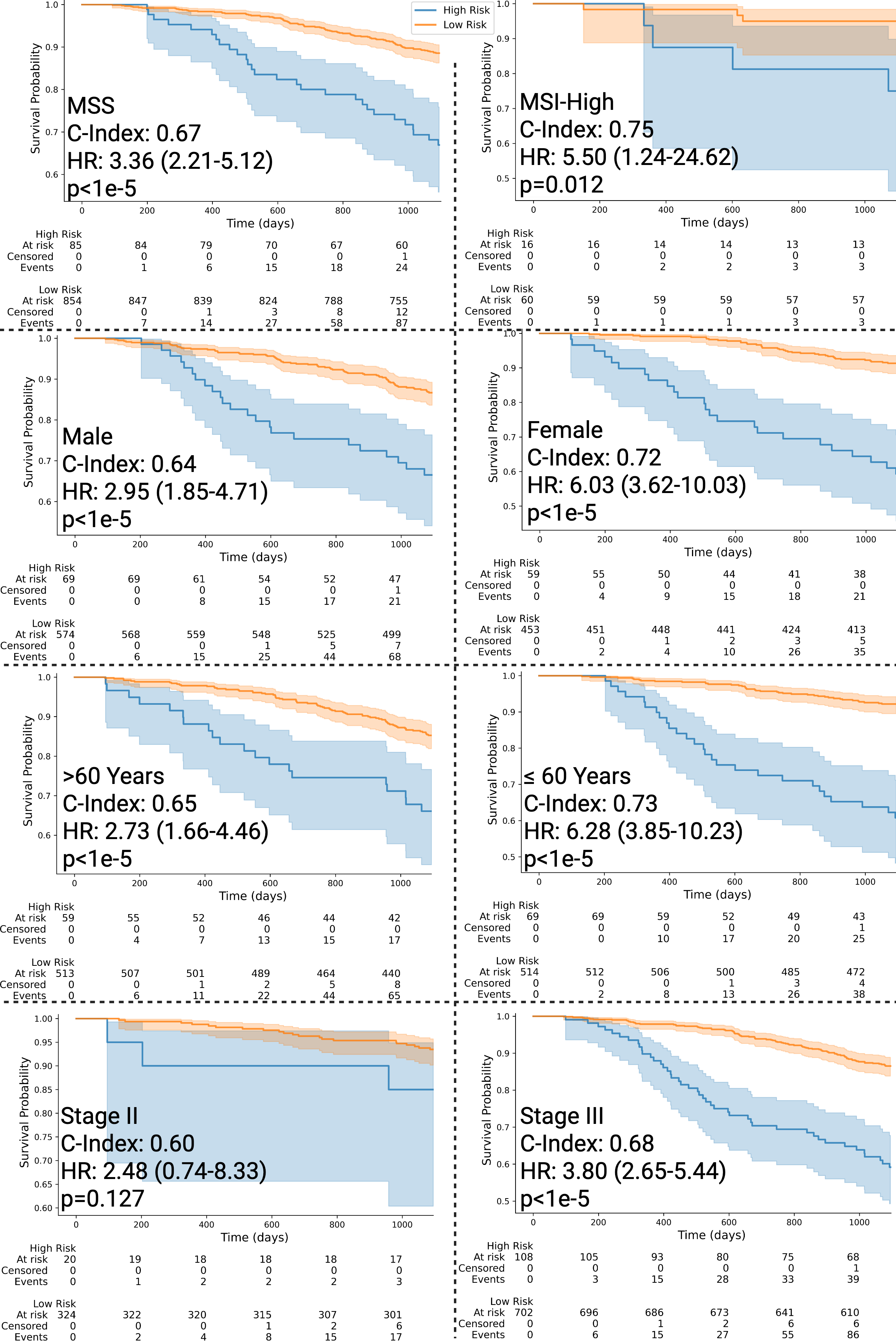}
    \caption{PETACC3 Kaplan-Meier Curves. For different clinical subpopulations, we show the performance of INSIGHT via a Kaplan-Meier Curve. Patients are stratified into high and low risk groups based on an optimal risk score threshold learned during training.}
    \label{fig:petacc_km}
\end{figure}

\begin{figure}
    \centering
    \includegraphics[width=0.7\linewidth]{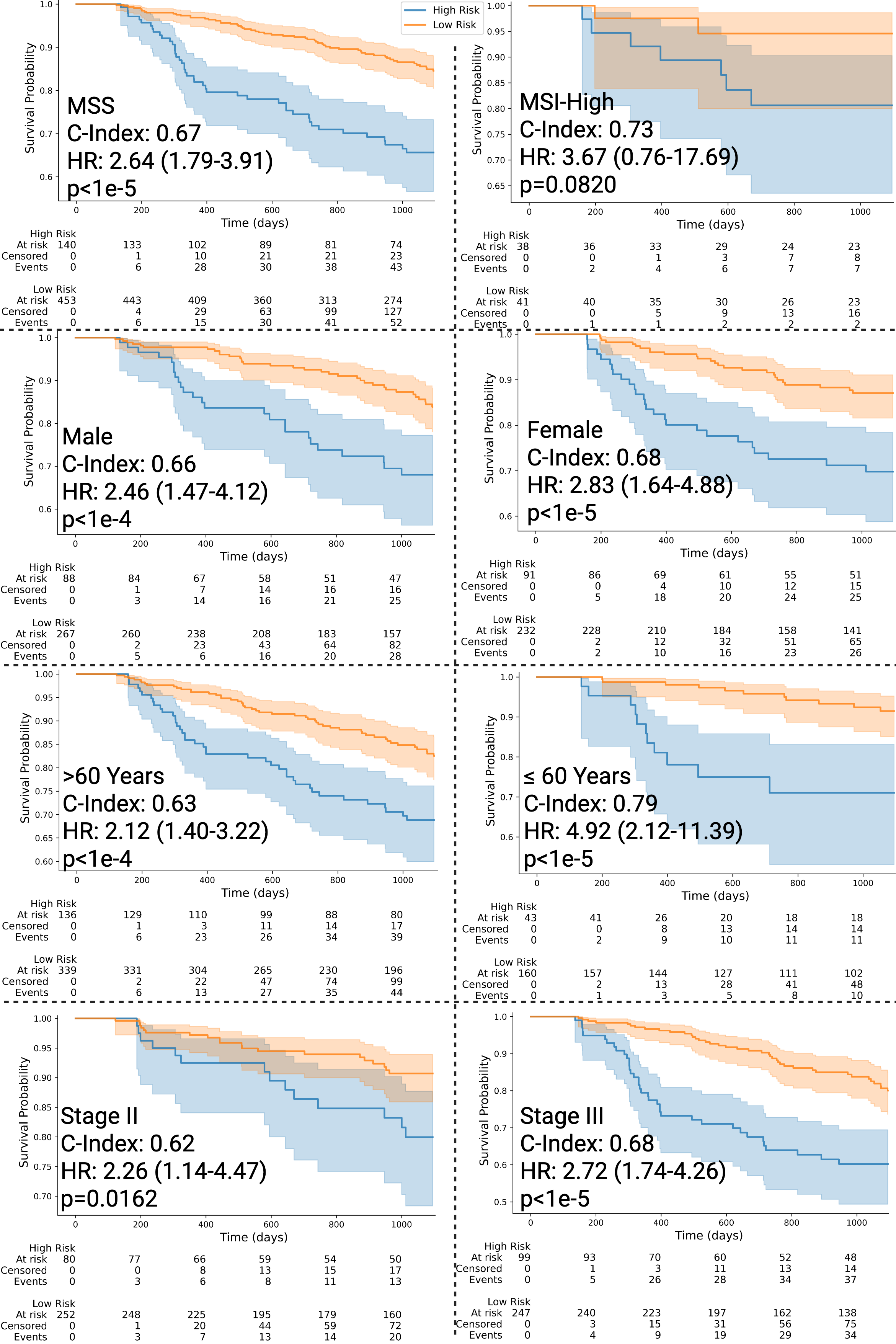}
    \caption{TCGA and SR386 Kaplan-Meier Curves. For each clinical subpopulation, we evaluate INSIGHT using Kaplan-Meier curves. Risk scores are obtained via 4-fold cross-validation: for each patient, we retain the risk score produced when that patient appeared in the test fold. This yields a combined dataset in which every patient is included exactly once and only as a test case. Patients are then stratified into high- and low-risk groups using an optimal risk score threshold learned during training.}
    \label{fig:TCGA_km}
\end{figure}

\begin{figure}
    \centering
    \includegraphics[width=0.8\linewidth]{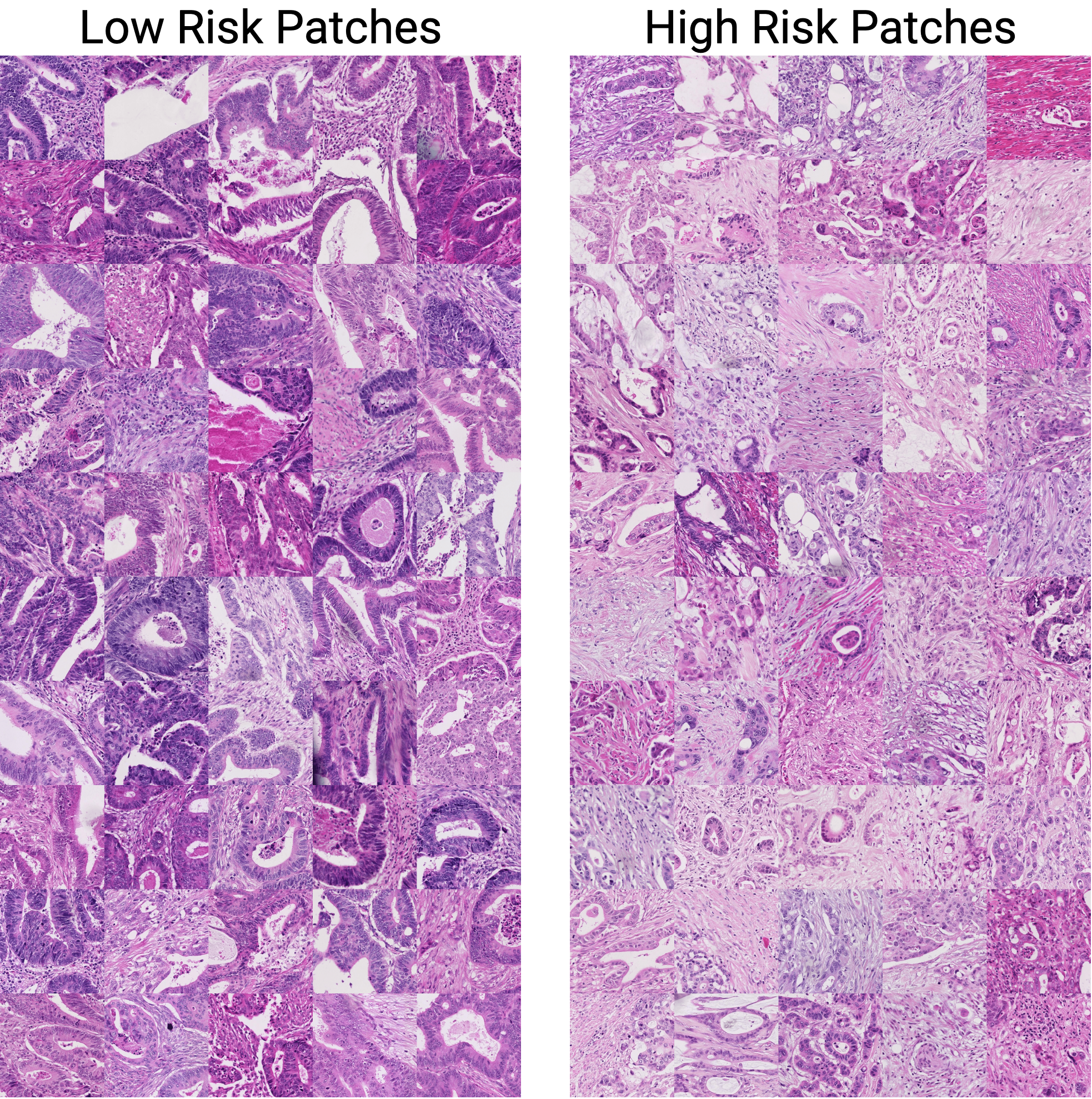}
    \caption{Representative high and low risk patches in PETACC3. For each PETACC-3 whole-slide image, we first identify patches classified as low risk (risk score < 0.3). From these, we select a single representative patch per slide by choosing the medoid patch, ensuring that each slide contributes equally and avoiding bias from multiple low-risk patches within the same WSI. We then perform k-medoids clustering (k = 50) on the resulting set of low-risk patches to identify distinct low-risk morphological themes across the cohort. A similar process was repeated for high-risk (risk score > 0.7) patches.}
    \label{fig:high_low_patches}
\end{figure}

\begin{figure}
    \centering
    \includegraphics[width=0.7\linewidth]{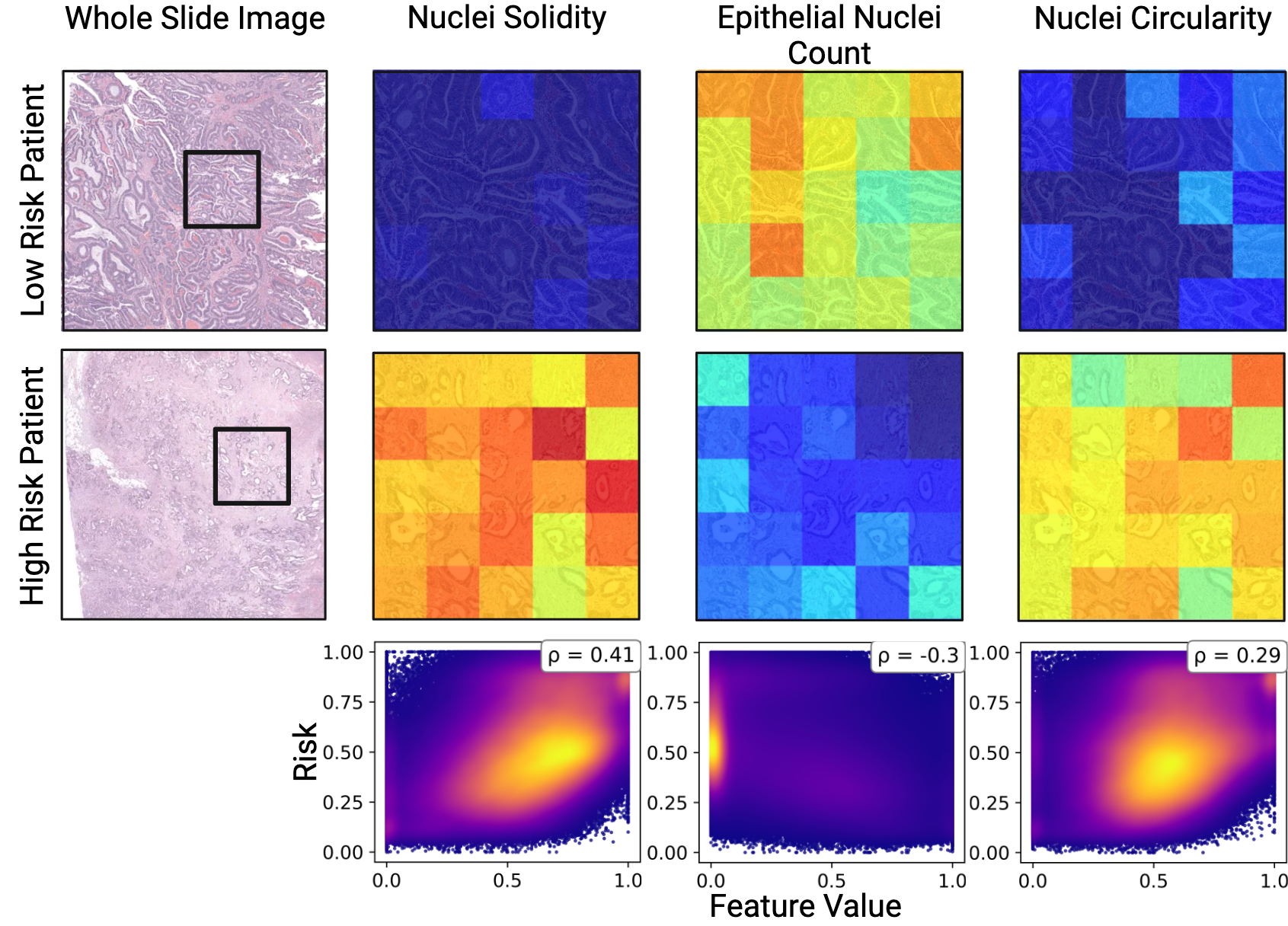}
    \caption{Example patches illustrating histomorphological features and their association with risk. We show two regions from whole-slide images in the Janssen cohort, one from a low-risk patient and one from a high-risk patient. For each region, we display the distribution of selected histomorphological features, including nuclear solidity and circularity. The high-risk region exhibits a greater proportion of patches with higher nuclear solidity and circularity. We also present a scatter-density plot of all patches in the cohort, showing patch-level risk scores versus the corresponding values of different histomorphological features, with colour indicating point density (blue to yellow denotes low to high density).}
    \label{fig:janssen_vis}
\end{figure}

\begin{figure}
    \centering
    \includegraphics[width=0.7\linewidth]{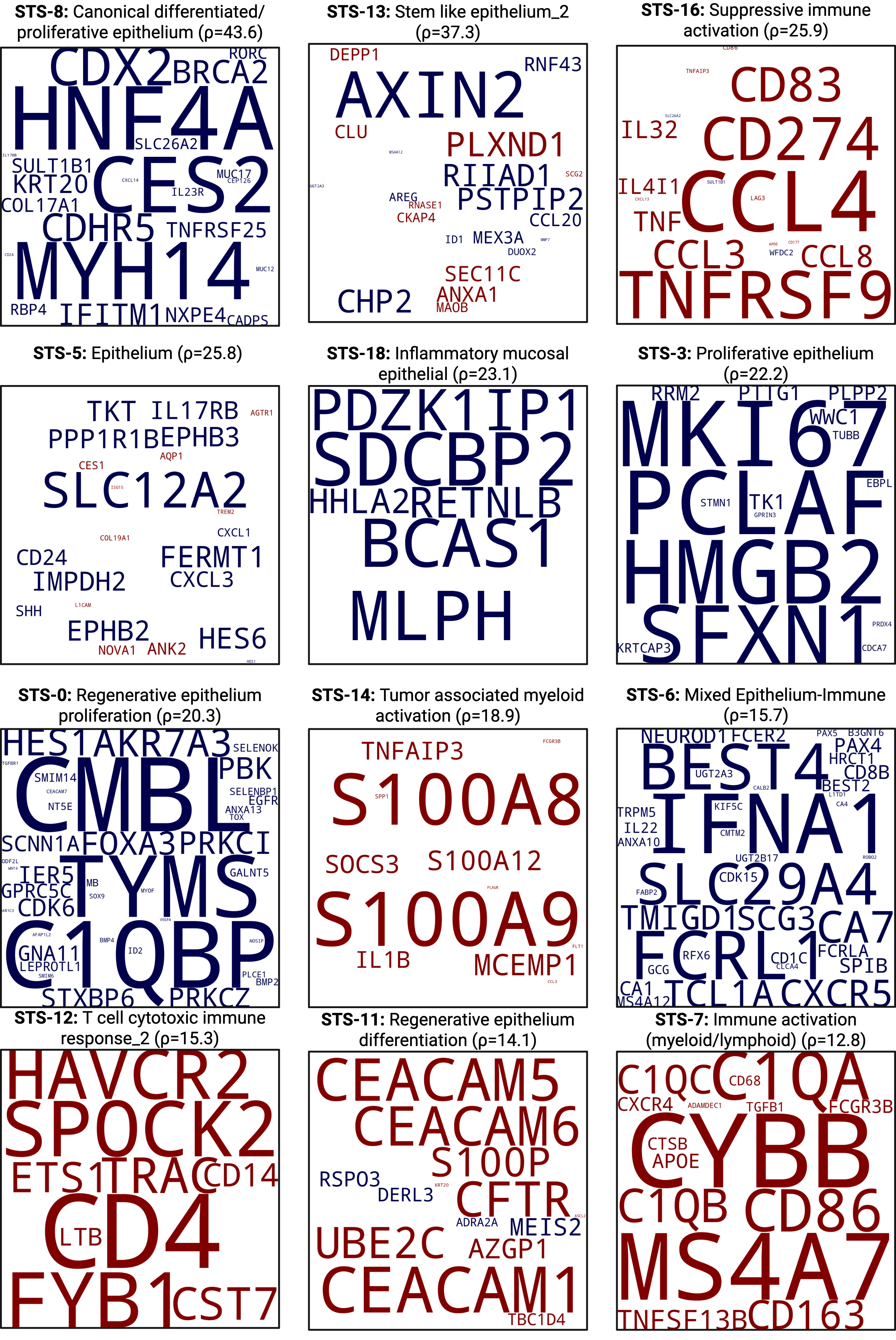}
    \caption{Spatial Transcriptomic Signature (STS) definitions. We present the first 12 spatial transcriptomic signatures (STS). Each signature is visualized as a word cloud of its defining genes. When a signature is active, genes shown in red are typically overexpressed, whereas genes shown in blue are underexpressed; the inverse pattern is observed when the signature is inactive. Gene size reflects its importance in defining the signature. We additionally report the univariate spearman correlation between each signature and localized patch-level risk.}
    \label{fig:topics_part_1}
\end{figure}

\begin{figure}
    \centering
    \includegraphics[width=0.7\linewidth]{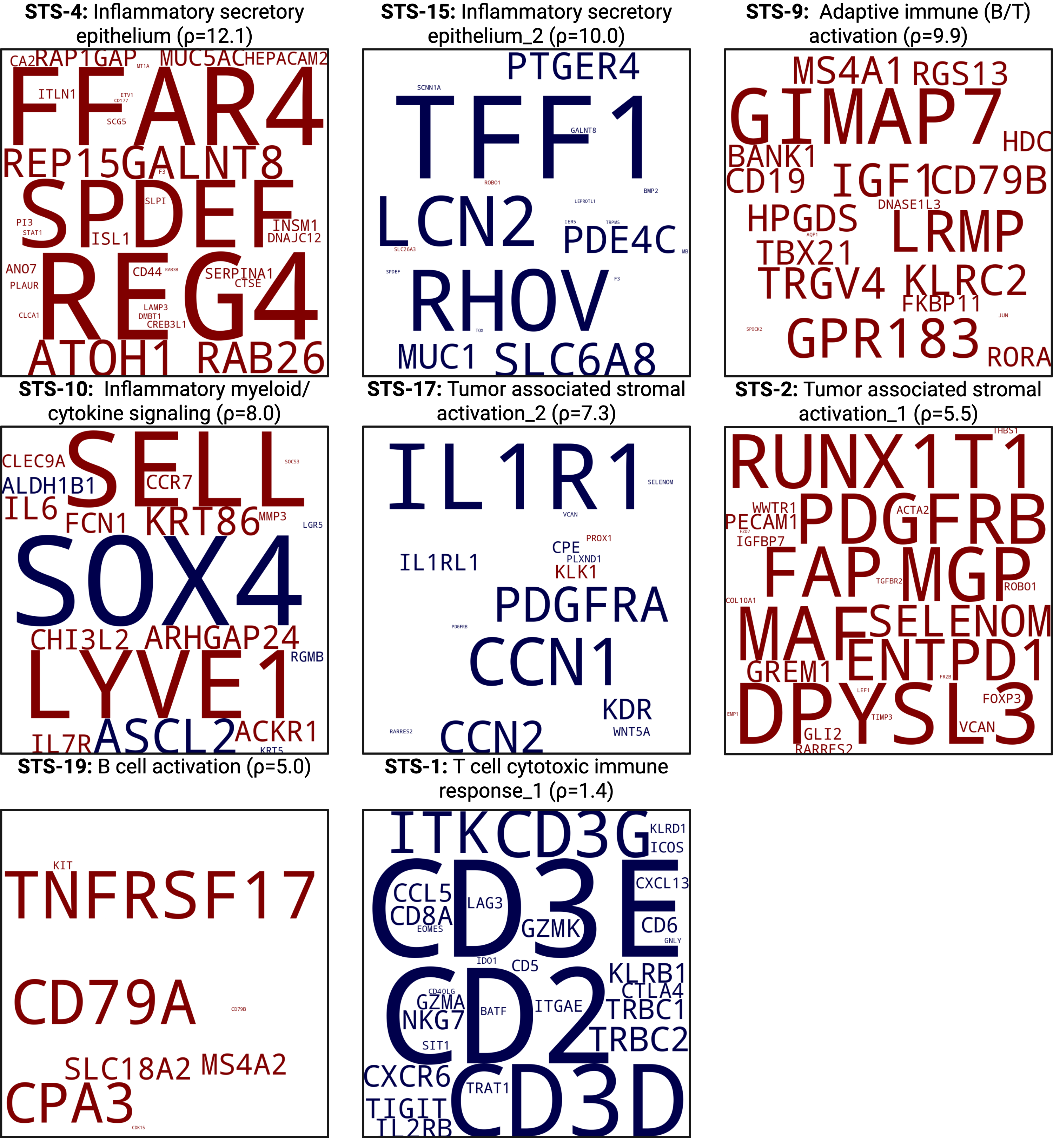}
    \caption{Spatial Transcriptomic Signature (STS) definitions. We present the last 8 spatial transcriptomic signatures (STS). Each signature is visualized as a word cloud of its defining genes. When a signature is active, genes shown in red are typically overexpressed, whereas genes shown in blue are underexpressed; the inverse pattern is observed when the signature is inactive. Gene size reflects its importance in defining the signature. We additionally report the univariate spearman correlation between each signature and localized patch-level risk.}
    \label{fig:topics_part_2}
\end{figure}

\begin{figure}
    \centering
    \includegraphics[width=0.7\linewidth]{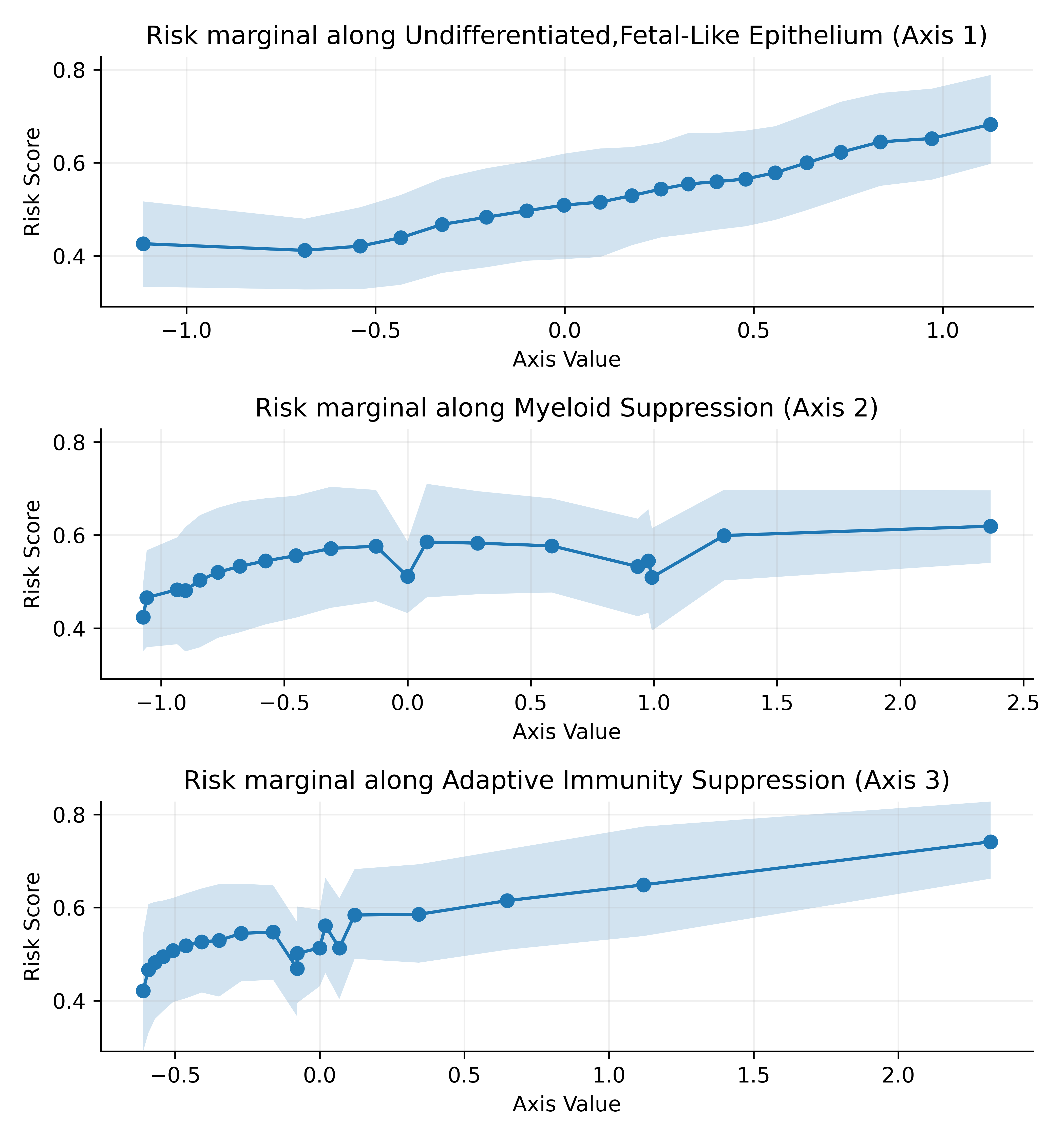}
    \caption{Risk Manifold Axes Marginals. We report the marginal effects of the three risk manifold axes, showing how increases in axis values (x-axis) are associated with changes in localized risk (y-axis).}
    \label{fig:manifold_marginals}
\end{figure}

\begin{figure}
    \centering
    \includegraphics[width=0.7\linewidth]{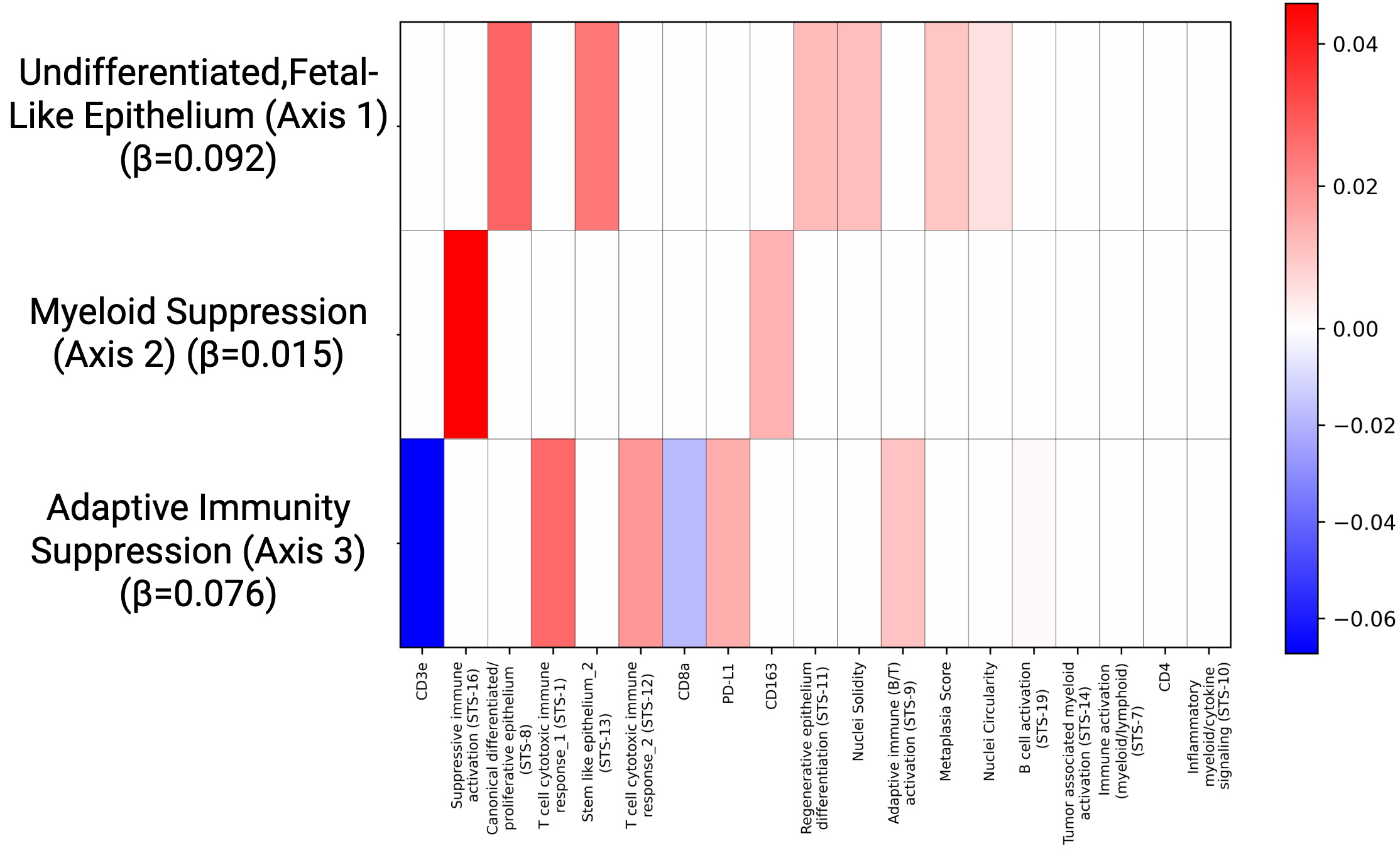}
    \caption{Risk Manifold Axes Feature Contributions. For each feature used in the manifold, we show its contribution to the definition of the final axis values. Feature weights are coloured by importance, with red indicating strong positive contributions and blue indicating strong negative contributions. Features shown in white have zero weight and therefore do not contribute to the corresponding axis. For each axis, we also report its overall importance via the associated $\beta$ coefficient}
    \label{fig:manifold_weights}
\end{figure}

\begin{table}[htbp]
\centering
\caption{STS OLS regression coefficients with FDR-adjusted significance. Using the 20 spatial transcriptomic signatures (STS), we fit an ordinary least squares (OLS) regression model to predict INSIGHT-derived localized risk. For each signature, we report the estimated regression coefficient ($\beta$), which indicates the direction and magnitude of the signature’s contribution to explaining risk. Statistical significance is assessed using false discovery rate (FDR) correction, and the corresponding FDR-adjusted q-value is reported.}
\label{tab:ols_results}
\begin{tabular}{lcc}
\hline
Spatial Transcriptomic Signature & Coefficient ($\beta$) & FDR $q$-value \\
\hline
Constant & 0.275 & 0.000 \\
STS-8  & 0.099 & 0.000 \\
STS-11 & 0.093 & 0.000 \\
STS-13 & 0.051 & $5.4 \times 10^{-111}$ \\
STS-5  & 0.041 & $4.4 \times 10^{-29}$ \\
STS-10 & $-0.038$ & $3.1 \times 10^{-44}$ \\
STS-16 & 0.038 & $8.1 \times 10^{-66}$ \\
STS-17 & 0.028 & $2.8 \times 10^{-32}$ \\
STS-1  & 0.025 & $1.6 \times 10^{-22}$ \\
STS-9  & 0.024 & $2.3 \times 10^{-20}$ \\
STS-14 & 0.023 & $5.4 \times 10^{-29}$ \\
STS-2  & 0.023 & $3.8 \times 10^{-21}$ \\
STS-18 & 0.014 & $5.4 \times 10^{-6}$ \\
STS-4  & 0.013 & $2.6 \times 10^{-7}$ \\
STS-12 & 0.011 & $4.3 \times 10^{-5}$ \\
STS-15 & $-0.011$ & $8.1 \times 10^{-5}$ \\
STS-0  & $-0.005$ & 0.169 \\
STS-19 & $-0.003$ & 0.141 \\
STS-3  & 0.003 & 0.440 \\
STS-6  & 0.001 & 0.524 \\
STS-7  & $-0.001$ & 0.740 \\
\hline
\end{tabular}
\begin{flushleft}
\footnotesize
Notes: Coefficients ($\beta$) estimated using ordinary least squares. $q$-values are false discovery rate-adjusted p-values using the Benjamini-Hochberg procedure.
\end{flushleft}
\end{table}

\end{document}